\documentclass[useAMS,usegraphicx]{mn2e}
\usepackage{lscape}
\usepackage[totalwidth=480pt,totalheight=680pt]{geometry}
\title[Jet SED from internal shocks]{The spectral energy distribution of compact jets powered by internal shocks}
\author[J. Malzac]{Julien Malzac$^{1,2}$\thanks{E-mail: julien.malzac@irap.omp.eu}\\
$^{1}$Universit\'e de Toulouse; UPS-OMP; IRAP;  Toulouse, France, \\ $^{2}$CNRS; IRAP; 9 Av. colonel Roche, BP 44346, F-31028 Toulouse cedex 4, France}
\begin{document}

\date{Accepted 2014 June 8,  Received 2014 May 11; in original form 2014 January 27.}

\pagerange{\pageref{firstpage}--\pageref{lastpage}} \pubyear{2008}

\maketitle

\label{firstpage}

\begin{abstract}

Internal shocks caused by fluctuations of the outflow velocity are likely to power the radio to IR emission of the compact jets of X-ray binaries.
The dynamics of internal shocks and the resulting spectral energy distribution (SED) of the jet are very sensitive to the time-scales and amplitudes of the velocity fluctuations injected at the base of the jet. I  present a new  code designed to simulate the synchrotron emission of a compact jet powered by internal shocks. I also develop a semi-analytical formalism allowing one to estimate the observed SED of the jet as a function of the Power Spectral Density (PSD) of the assumed fluctuations of the Lorentz factor.  I discuss the cases of a sine modulation of the Lorentz factor and Lorentz factor fluctuations with a power-law PSD shape.  Independently of the details of the model,  the observed nearly flat  SEDs are obtained for PSDs of  Lorentz factor  fluctuations that are close to a flicker noise spectrum (i.e.  $P(f)\propto1/f$). The model also presents a strong wavelength dependent variability that is similar to that observed in these sources.

\end{abstract}


\section{Introduction}

Compact jets of X-ray binaries have an approximatively flat spectral energy distribution (SED) extending from the radio to the mid-IR (e.g.  Corbel et al. 2000; Fender et al. 2000; Fender 2001; Corbel \& Fender  2002). These flat, or slightly inverted, spectra are usually ascribed to self-absorbed synchrotron emission from conical compact jets (Condon \& Dressel 1973; de Bruyn 1976; Marscher 1977;  Konigl 1981; Ghisellini, Maraschi \& Treves 1985) under the assumption of continuous energy replenishment of the adiabatic losses  (Blandford \& K\"onigl 1979).  The compensation of these energy losses is crucial for maintaining this specific spectral shape. In absence of any additional dissipation mechanism along the jet, the flows quickly cools down as it expands and this leads to a strongly inverted radio spectrum (Marscher 1980). 

This problem can be avoided by assuming a non-conical jet, but  fine tuning of the jet geometry is required in order to produce the observed flat SEDs (Hjellming \& Johnston 1988; Kaiser 2006).
There is also the possibility that X-ray binary jets could be unconfined,  then adiabatic losses would be strongly reduced. Falcke (1996,  see also  Falcke \& Markoff  2000) has proposed a freely expanding jet model with negligible radial expansion losses. However the observational evidence indicates that jets are indeed confined and if this was not the case, the current observational constraints on the jet opening angle in X-ray binaries requires jet Lorentz factors greater than $\sim$10 (Miller-Jones, Fender \& Nakar 2006). That is much larger than usually believed and also greater than the jet Lorentz factor predicted by the model of Falcke (1996). 

This suggests that some form of dissipation occurring at large distance from the compact object does compensates for the adiabatic losses. In conical jets the dissipation rate must be distributed along the jet and scale like $z^{-1}$ in order to reproduce the flat observed SED (Malzac 2013, hereafter M13). In phenomenogical jet emission models such a dissipation process is postulated (Blandford \& Koenigl 1979; Zdziarski, Pjanka \& Sikora, 2013) but  the physical origin of this  $\propto z^{-1}$ dissipation profile is not treated. Such a dissipation could occur through conversion of either the jet Pointing flux, or the jet kinetic energy into internal energy. Lyubarsky (2010)  suggested a mechanism of conversion of Pointing flux into internal energy through magnetic reconnection. The dissipation profile expected from such processes is however difficult to predict and currently unknown.

In this paper I will instead explore the second possibility, and consider the conversion of jet kinetic energy through internal shocks. Internal shocks models have been proposed to explain the multi-wavelength emission of jetted sources such as $\gamma$-ray bursts (Rees \& Meszaros 1994; Daigne \& Moscovitch 1998),  active galactic nuclei (Rees 1978; Spada et al. 2001; B{\"o}ttcher 
\& Dermer 2010), or microquasars (Kaiser, Sunyaev \& Spruit 2000; Jamil, Fender \& Kaiser  2010, JFK10 hereafter).
In these models, shells of gas are continuously ejected with randomly variable velocities and then propagate along the jet. At some point, the fastest fluctuations start catching up and merging with slower ones. This leads to the formation of shocks in which electrons are accelerated up to relativistic energies. These electrons then produce synchrotron and also, possibly, inverse Compton radiation.
 
 A point that has been generally overlooked in all these studies is that the result is very sensitive to the properties  of the assumed fluctuations of the Lorentz factor of the jet. 
Recently, M13 showed that the adiabatic expansion losses can be entirely compensated by internal shocks, provided that the fluctuations of the jet Lorentz factor correspond to a flicker noise process i.e. with the Fourier Power Spectral Density (PSD), $S(f)$, of the fluctuations which is inversely proportional to the Fourier frequency $f$. 

The aim of this paper is to explore the sensitivity of the jet SED to the shape of the PSD of the fluctuations as well as to the details of the modelling of the internal shocks. 
In Section~\ref{sec:analytic}, I generalise the analytical formalism of M13 to calculate the shock  dissipation profile in the jet and its resulting time-averaged SED for input fluctuations with any PSD shape. In Section~\ref{sec:numeric}, I describe a new numerical code designed to simulate the collisions and shock dissipation in the ejecta. I also investigate the sensitivity  of the results to various approximations in the modelling of the shell collisions.   
 In Section~\ref{sec:results}, I use both the analytical estimates and simulations to study the response of the jet to sinusoidal fluctuations, and fluctuations with a power-law PSD, I also demonstrate the capability of the code in predicting the variability properties of the sources. 

\section{Analytical Model}\label{sec:analytic}

\subsection{Hierarchical damping of the fluctuations}

Consider a jet of time-average bulk Lorentz factor $\gamma$.
The jet is modelled as a set of discrete homogeneous ejecta that are ejected at regular time intervals $\Delta t $ comparable to the dynamical time-scale of the accretion flow (i.e a few ms). I assume small, time dependent  fluctuations $\Delta\gamma$  of the Lorentz factor so that the shell of gas ejected at time $t$ at the base of the jet has a Lorentz factor $\gamma+\Delta\gamma(t)$. Those ejecta then propagate ballistically along the jet axis. For simplicity, in these analytical estimates, the mass of the ejecta is assumed to be a constant $m_0=0.5P_j\Delta t/(\gamma-1)c^2$, where $P_j$  is the total kinetic power of the two-sided jets (hence the factor 0.5) and $c$ the speed of light. The effects of a varying mass of the ejecta will be investigated numerically in Section~\ref{sec:shellinjection}, they are usually weak. 

The Fourier transform of the fluctuations is defined as: 
\begin{equation} 
\Delta\Gamma(f)=\int_{-\infty}^{+\infty} \Delta\gamma(t) e^{-2i\pi f t} dt
\end{equation}
Since the Lorentz factor does not vary on time scales shorter than $\Delta t$, there is a Fourier frequency  $f_0=1/(2\Delta_t)$ above which $\Delta\Gamma(f>f_0)=0$. In other words $f_0$ is the highest possible frequency of the injected fluctuations. 
Therefore, the initial variance of the injected fluctuations can be written as:
\begin{equation}
\gamma_{\rm rms0}^2=\int_{0}^{f_0}2S(f)df,
\end{equation}
where $S=|\Delta\Gamma|^2$ is the  Fourier Power Spectral Density (PSD) of the fluctuations.  

On average  the centre of momentum of two neighbouring shells at injection are separated by a distance $\lambda_0=\beta c \Delta t$ (where $\beta =\sqrt{1-\gamma^{-2}}$ is the ratio of the jet velocity to the speed of light). This length also corresponds to the smallest scale of the velocity fluctuations.  
As the ejecta propagate downstream along the $z$ direction, the fastest shells catch up and merge with slower ones. During this process of hierarchical merging the average mass of the ejecta and their separation will increase $\lambda(z)/\lambda_0=m(z)/m_0=K(z)$. 
This growth in length scale implies a damping of the fluctuations of frequencies higher than $f_0/K$. As a consequence, the variance of the fluctuations decreases: 

\begin{equation}
\gamma_{\rm rms}^2(K)=\int_{0}^{\frac{f_0}{K}}2S(f)df.
\label{eq:gammadek}
\end{equation}

The problem appears simpler when viewed in the frame moving with a Lorentz factor $\gamma$. In the limit of small scales fluctuations, this frame coincides with the frame of the centre of momentum of the shells. To first order in $\gamma_{\rm rms}/\gamma$, the average velocity of the shells is 0 and its rms amplitude:
\begin{equation}
\tilde{v}_{\rm rms}(K)=\frac{\gamma_{\rm rms}(K)c}{\gamma \beta}.
\end{equation}

In the limit of low amplitude fluctuations, the velocities in the moving frame are non-relativistic. 
All of the kinetic energy of the shells is available for conversion into internal energy. After all the fluctuations have been damped, the flow will have received  a total energy $\tilde{u_0}=\tilde{v}^2_{\rm  rms}(K=1)/2$ per unit mass. The free specific energy of the system (i.e available for dissipation) is:
\begin{equation}
\tilde{u}(K)=\frac{\tilde{v}^2_{\rm rms}(K)}{2}=\frac{\gamma^{2}_{\rm rms}(K)c^2}{2\gamma^2\beta^2},
\end{equation}
then, from equation~(\ref{eq:gammadek}):
\begin{equation}
\frac{d\tilde{u}}{dK}=-\frac{f_0 S(f_0/K) c^2}{K^2\gamma^2\beta^2}.
\label{eq:dusdk}
\end{equation}

The fluctuations of scale $K$ collide and merge after travelling for a time: 
\begin{equation}
\tilde{t}(K)=y\tilde{\lambda}/\Delta\tilde{v}=\frac{y\gamma^2\beta^2 K}{4f_0\sqrt{J(K)}},
\label{eq:tdek}
\end{equation}
where $y$ is a factor of the order of unity accounting for  the effects of the longitudinal extension of the ejecta. It will be estimated in Section~\ref{sec:fc}. $\Delta \tilde{v}$ is the average (absolute) velocity difference between the merging shells. In the case of a linear variability process it can be estimated as:
\begin{equation}
{\Delta{\tilde{v}}}^2=\left(\frac{ 2c}{\gamma\beta }\right)^2 J(K),
\label{eq:dv2}
\end{equation}
with
\begin{equation}
J(K)=\int_0^{f_0/K}S(f)\left[1-\cos{(K \pi f/f_0)}\right]df.
\label{eq:jk}
\end{equation}

Let $\tilde{\epsilon}$ be the specific internal energy of the flow. Its evolution is governed by:
\begin{equation}
\frac{d\tilde{\epsilon}}{dK}=-\frac{d{\tilde{u}}}{dK}-(\gamma_a-1)\tilde{\epsilon}\frac{ d\ln \tilde{V}}{dK},
\label{eq:desdk}
\end{equation}
where $\tilde{V}$ is the comoving volume.  The first term on the right-hand side of equation~(\ref{eq:desdk}) is energy injection through internal shocks, the second term accounts for adiabatic losses (see e.g. equation (1) of M13). $\gamma_a$ is the effective adiabatic index of the flow. For simplicity, it is assumed here that all the components of the flow (thermal particles, relativistic particles, magnetic field...)  have the same adiabatic index corresponding to that  of a relativistic gas (i.e. $\gamma_a=4/3$).

Neglecting the possible longitudinal expansion losses,  the formal solution is:
\begin{equation}
\tilde{\epsilon}=\tilde{\epsilon_{0}}\left[\frac{R_b}{R(K)}\right]^{2\gamma_a-2}-\int_1^{K}\frac{d\tilde{u}}{dK'}\left[\frac{R(K')}{R(K)}\right]^{2\gamma_a-2} dK',
\label{eq:evole}
\end{equation}
where  $R$ is the jet radius after a time $\tilde{t}$. The function $R(K)$ defines the jet geometry.
For simplicity, in this paper we will consider only conical jets:  
\begin{equation}
R=R_b+ \tilde{t}(K) \gamma\beta c \tan{\phi}.
\label{eq:rdet}
\end{equation}
where $\phi$ is the jet half opening angle and $R_b$ the radius of the jet at its base.

For a specific PSD of the Lorentz factor fluctuations, the specific energy profile $\tilde{\epsilon}(z)$ can be evaluated by combining equation~(\ref{eq:evole}) with equations~(\ref{eq:dusdk}),  (\ref{eq:tdek}) and (\ref{eq:rdet}).

A fraction of the internal energy is in the form of relativistic electrons. Another fraction is in the thermal energy of protons, the remaining fraction is in the form of magnetic field energy. 
{The magnetic field energy density is:
\begin{equation}
\frac{B^2}{8\pi}=\frac{\tilde{\rho}\tilde{\epsilon}}{1+\xi_{\rm e}+\xi_{\rm p}},
\end{equation}
where $\xi_{\rm p}$ and $\xi_{\rm e}$ are  the equipartition factors related to the thermal energy of the ions $U_{\rm p}=\xi_{\rm p}B^2/8\pi$, and  to the internal energy of the accelerated electrons $U_{\rm e}=\xi_{\rm e}B^2/8\pi$ respectively.  $\xi_{\rm e}$ and $\xi_{\rm p}$ are assumed to be constant and uniform along the jet. Once the jet geometry is fixed, the average mass density $\tilde{\rho}$  of a shell with radius $R$ is given by:
\begin{equation}
\tilde{\rho}=\frac{P_{J} R^{-2}}{2f_{\rm v}(\gamma-1)c^2 \gamma\beta c \pi }.
\label{eq:rho}
\end{equation}
where $f_{\rm v}$ is the volume filling factor of the shells. The volume filling factor of the colliding shells, $f_{\rm v}$, is defined so that the average length of a shell of separation $\lambda$ is $f_{\rm v}\lambda$. If $f_{\rm v}\ll1$, the jet is constituted of thin colliding `pancakes', while for $f_{\rm v}=1$, the jet is a continuous flow.  In principle, $f_{\rm v}$ depends on the properties of the shells at injection. But $f_{\rm v}$ is also expected to depend on $z$ (or $\tilde{t}$). Indeed, the shells are strongly compressed during the collisions,  then as they gain internal energy  they can also expand under the effect of their own pressure. Those effects are fully accounted in the numerical code, while a simple analytical treatment of the dynamic of the shells expansion and compression is proposed in Section~\ref{sec:fc}). The SED is not very sensitive to $f_{\rm v}$ and, for the purpose of finding analytical estimates of the SED,  $f_{\rm v}$ is taken to be a constant along the jet,  given by equation~(\ref{eq:fe}).

 The electron energy density is:
\begin{equation}
U_{\rm e}=\tilde{n}_{\rm e} \bar{\gamma}_{\rm e} m_e c^2 =\xi_{e} B^2/8\pi,
\label{eq:equipart}
\end{equation}
 where $\tilde{n}_e$ and $\bar{\gamma}_{\rm e}$ are the density and average Lorentz factors of the accelerated electrons. For simplicity I assume that the accelerated electrons have  a power-law particle distribution $n(\gamma)=n_0\gamma^{-p}$ in the range $\gamma_{\rm min}-\gamma_{\rm max}$ and that the shape of this distribution does not depend on  time or position in the jet. Its normalization varies to account for the energy gains and losses of the electrons. The average Lorentz factor of the electrons,
\begin{equation}
\bar{\gamma}_{\rm e}=\int_{\gamma_{\rm min}}^{\gamma_{\rm max}}\gamma^{1-p}d\gamma /\int_{\gamma_{\rm min}}^{\gamma_{\rm max}}\gamma^{-p}d\gamma,
\end{equation}
 is therefore a constant, and only the number of accelerated electrons varies. A different scenario in which the number of accelerated electrons is kept constant but $\bar{\gamma}_{\rm e}$ is allowed to vary is discussed in Appendix~\ref{sec:cnae} and found to give similar results.}

Once the average magnetic field, internal energy and density profile of the ejecta are determined, the time-averaged synchrotron emission of the jet can be estimated using standard formulae that are  detailed in Appendix~\ref{sec:analyticem}.

\subsection{Shell expansion and compression}\label{sec:fc}

In this section, I derive simple analytical estimates of the jet volume filling factor. 
 Here, as it is customary, I assume that this expansion occurs in the co-moving frame at a speed:
\begin{equation}
\tilde{v}_{\rm s}=\sqrt{\tilde{\epsilon}(\gamma_a-1)}=\Delta \tilde{v}/\mathcal{M}.
\end{equation}
As long as the gas is close to equipartition $\tilde{v}_{\rm s}$ is close to the sound speed and therefore $\mathcal{M}$ corresponds approximately to the mach number of the colliding shells. 

 For simplicity, I consider that the longitudinal expansion is free and therefore the adiabatic losses due to longitudinal expansion can be neglected. Any pressure work done during this expansion can only be used to accelerate or compress neighbouring shells so that overall, this does not represent  a loss of energy for the flow. I thus do not expect that taking these effects into account would change the results dramatically. 

 The shells can also undergo significant compression during shocks. When two shells are in the process of merging  the compression velocity is simply the difference of velocity of the two merging ejecta $\Delta \tilde{v}$. During a collision however, the compression lasts only for the time necessary for the shock to cross the shell, that is:
 \begin{equation}
  \tilde{t}_{\rm c}\simeq x f_{\rm v}\tilde{\lambda}/\Delta \tilde{v},
  \label{eq:tc}
  \end{equation}
 where $x^{-1}\Delta \tilde{v}$ is the speed of the shock in the frame of the unshocked ejecta. Applying the jump conditions of Blandford \& McKee (1976) and to first order in $\Delta \tilde{v}$, their equation (5) gives:
 \begin{equation}
 x^{-1} \simeq \sqrt{1/4+\gamma_a/2+(\gamma_a/2)^2}=7/6.
 \end{equation}
 During an infinitesimal time-interval $d\tilde{t}$, the length of a shell $\tilde{l}=f_{\rm v}\tilde{\lambda}$ changes by an amount: 
\begin{equation}
d\tilde{l}=f_{\rm v}d\tilde{\lambda}+df_{\rm v}\tilde{\lambda}.
\end{equation}
The fist term represents the increase in size of the ejecta due to the mergers. The second terms is the contribution from compression and expansion which can also be written as:
\begin{equation}
df_{\rm v}\tilde{\lambda}=2\tilde{v}_{s}d\tilde{t}-\Delta\tilde{v} d\tilde{t}_{\rm c},
\end{equation}
where $dt_c$ is the amount of time during which the shell was compressed in the interval $dt$. 
Or, equivalently:
\begin{equation}
\frac{df_{\rm v}}{dt}=\left(\frac{2}{\mathcal{M}}-\frac{dt_{c}}{dt}\right)\frac{\Delta\tilde{v}}{\tilde{\lambda}}.
\label{eq:dfsdt}
\end{equation}
For simplicity, I approximate the compression factor as: 
\begin{equation}
\frac{d\tilde{t}_{c}}{d\tilde{t}}\simeq\frac{\tilde{t}_{c}}{\tilde{t}} = \frac{x}{y}f_{\rm v},
\label{eq:dtcsdt}
\end{equation}
Then the $y$   parameter in equations~(\ref{eq:tdek}) and~(\ref{eq:dtcsdt}) is approximated as: 
\begin{equation}
y=\frac{1-f_{\rm v}}{1+2\mathcal{M}^{-1}}.
\label{eq:y}
\end{equation}
The numerator on the right hand side of equation~(\ref{eq:y}) accounts for the reduced distance between shells due to their longitudinal extension. The denominator corrects for the increased collision rate caused by the expansion velocity of the shells. Using this expression for $y$, equation~(\ref{eq:dfsdt}) can then be rewritten as: 
\begin{equation}
\frac{df_{\rm v}}{dt}=\left(\frac{1-f/f_e}{1+\mathcal{M}/2} \right) t^{-1},
\label{eq:dfsdt2}
\end{equation}
where
 \begin{equation}
 f_{\rm e}=\left[1+x\left(1+\mathcal{M}/2\right)\right]^{-1}.
 \label{eq:fe}
 \end{equation}
 In the case where the mach number is a constant, it  can be integrated easily and one gets: 
 \begin{equation}
 f_{\rm v}=f_e\left[1-\left(1-\frac{f_i}{f_e}\right)\left(\frac{t}{t_i}\right)^{-\frac{1+M/2}{f_e}}\right],
 \label{eq:fvdet}
 \end{equation}
where $f_{\rm i}$ is the initial value of the volume filling factor at time $t_{\rm i}$ when the Mach number became a constant.
We can see from equation~(\ref{eq:fvdet}) that at large times the system evolves asymptotically toward an equilibrium in which  $f_{\rm v}=f_{\rm e}$. As can be seen from  equations (\ref{eq:dfsdt2}) and  (\ref{eq:fvdet}),  the changes in volume factor are driven by the variations of $\mathcal{M}$. Therefore, if the variations of the Mach number are negligible the volume filling factor tends toward a constant $f_{\rm e}$ given by equation~(\ref{eq:fe}). In fact, the simulations show that the evolution of $f_{\rm v}$ is more complex, but for the cases investigated in this paper it turns out that assuming a constant $f_{\rm v}=f_e$ leads to a good approximation of the SEDs.
 
 \begin{figure}
\includegraphics[width=\linewidth]{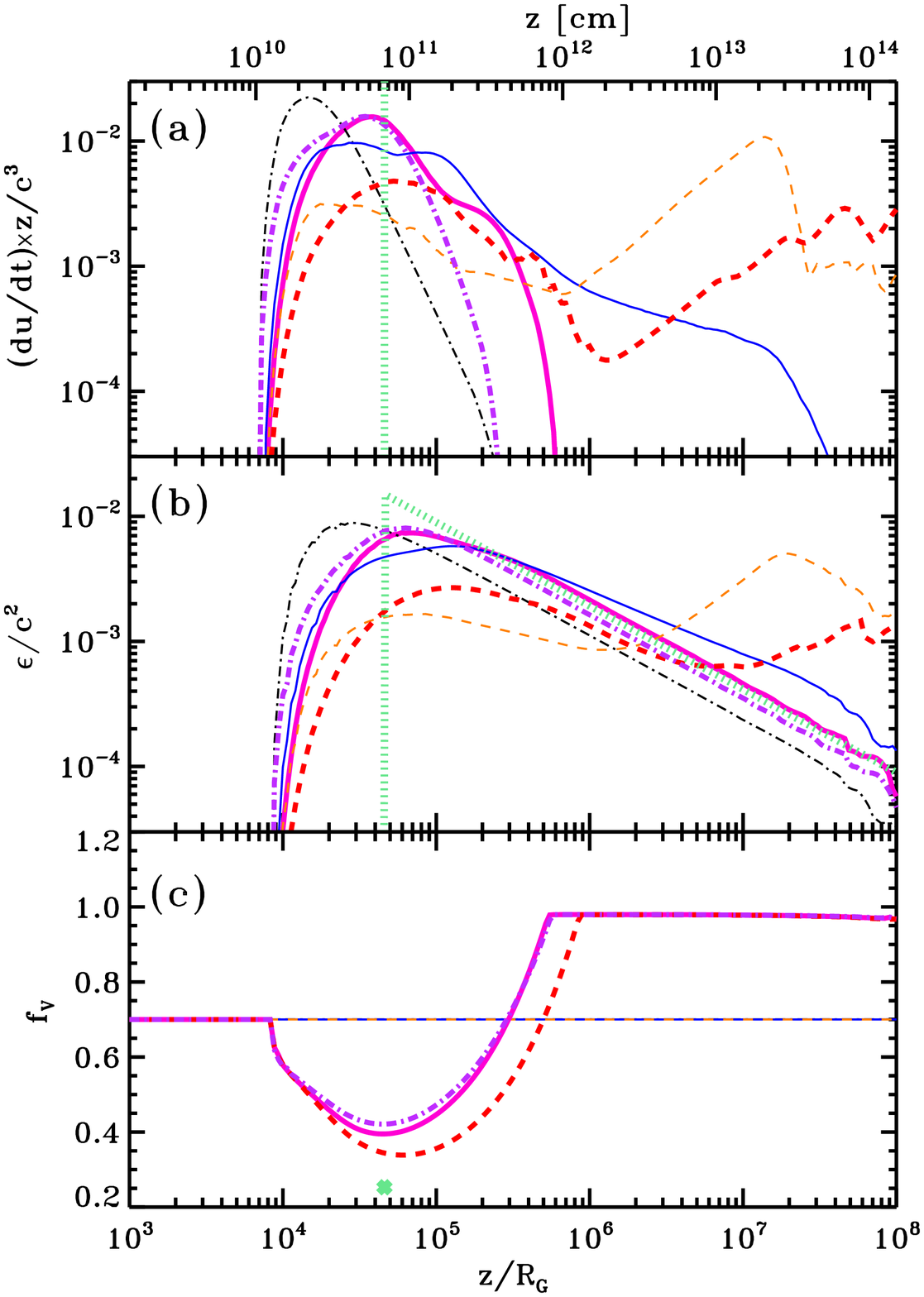}
\includegraphics[width=\linewidth]{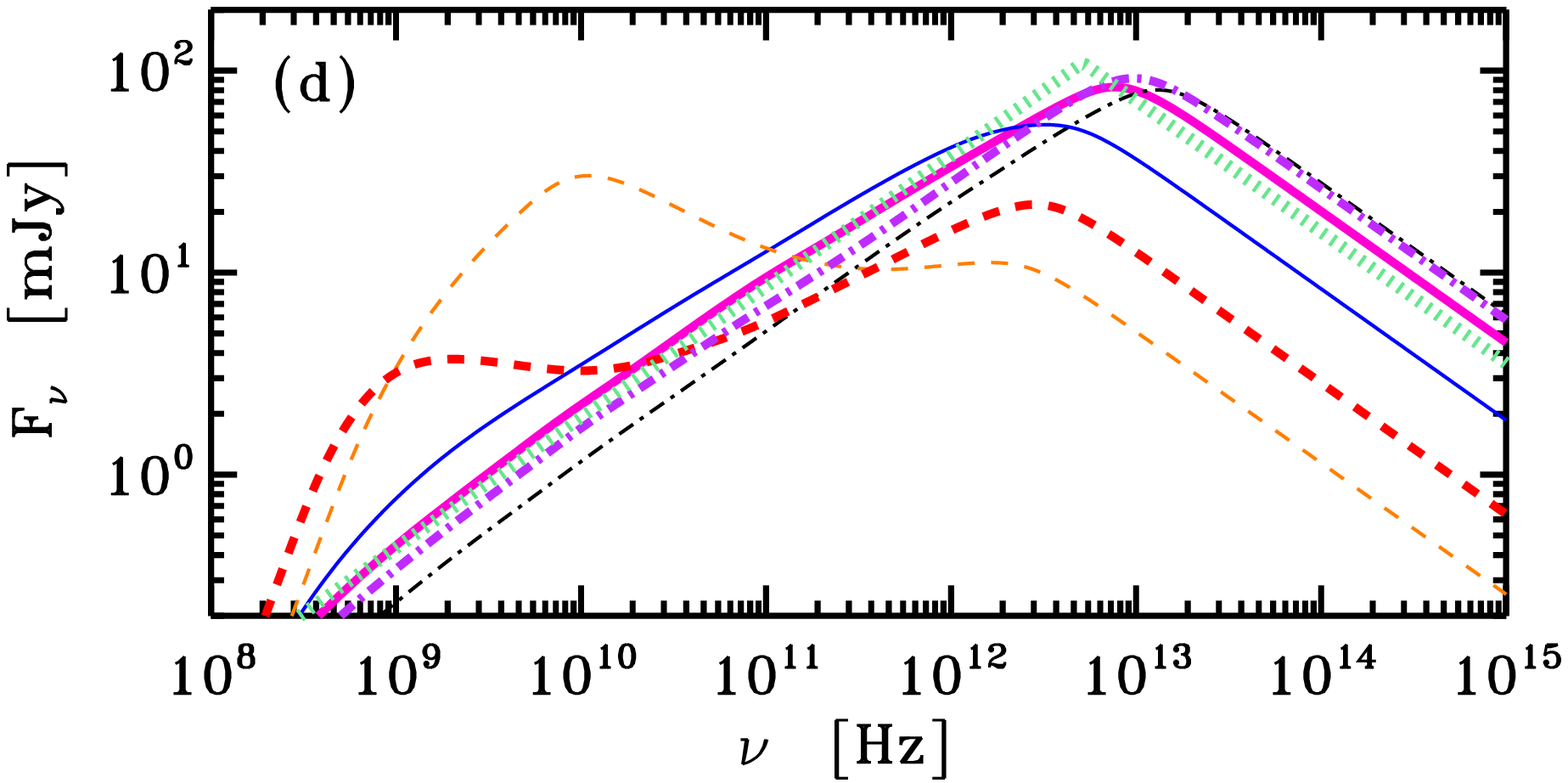}
 \caption{Simulation of the internal shock model assuming sinusoidal fluctuations of the jet Lorentz factor of frequency 1 Hz  and rms amplitude $\gamma_{\rm rms}=0.3$. Panel (a): time averaged shock energy dissipation rate along the jet as a function of distance from the base. Panel (b): time-averaged specific energy profile along the jet. Panel (c): average volume filling factor of the shells.  Panel (d): predicted time averaged SEDs calculated for an inclination angle of 40 degrees and a distance to the source of 2 kpc. The emission from the counter-jet is not shown.  In all panels, the different types of curves stand for different assumption regarding  shock dissipation. The thick and thiner curves show the result of simulations obtained with standard and solid shell assumptions respectively. The full curves show the results for `fast' dissipation while the dashed curves show the results for slow dissipation, and dash-dotted curves are for instantaneous dissipation at the time of contact.  The dotted curve shows the results expected from the analytical estimates. See Sections~\ref{sec:shellcoll} and~\ref{sec:sin} for details.
  }
 \label{fig:sin}
 \end{figure}

\section{Numerical model}\label{sec:numeric}

\begin{figure}
\includegraphics[width=\linewidth]{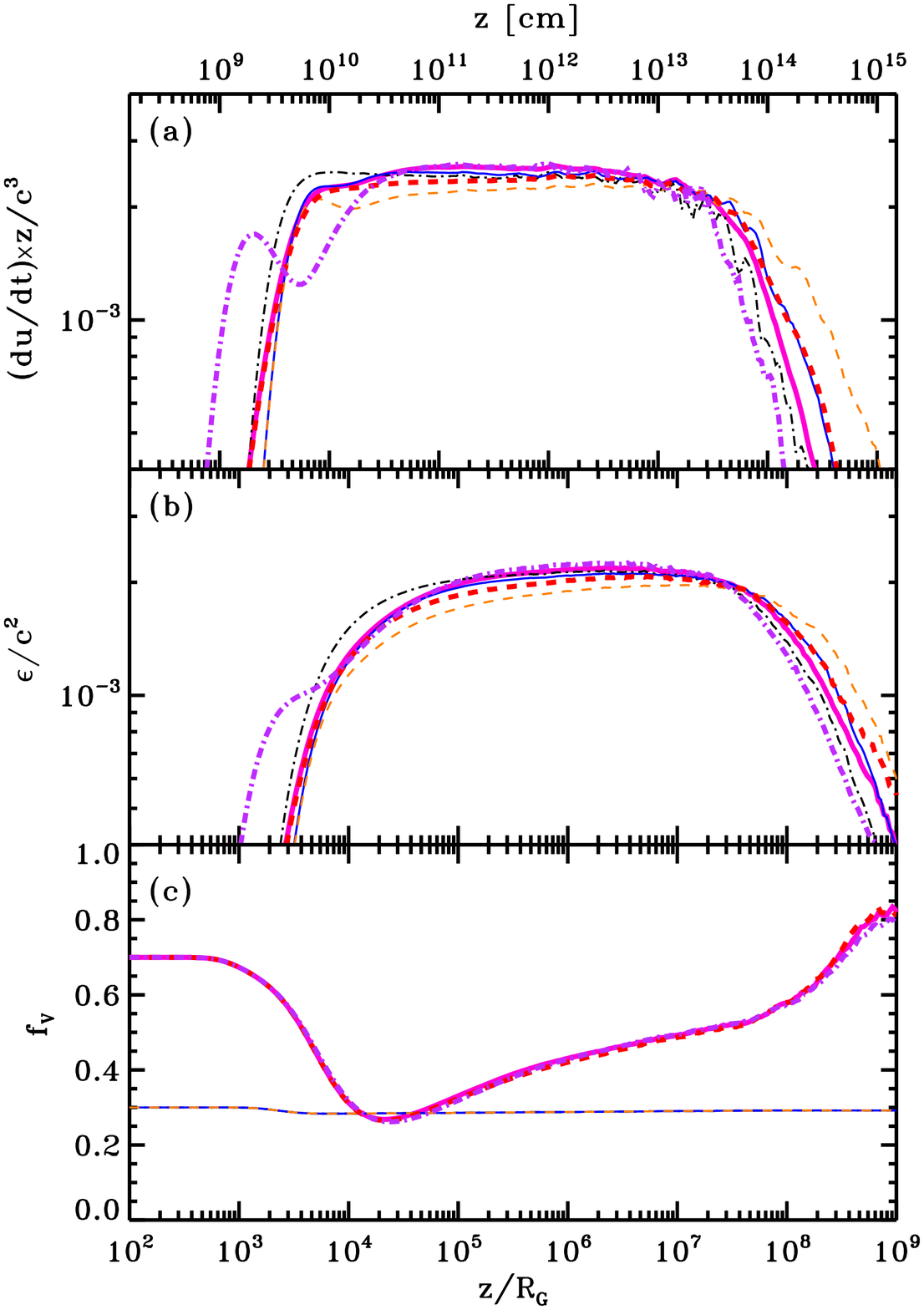}
\includegraphics[width=\linewidth]{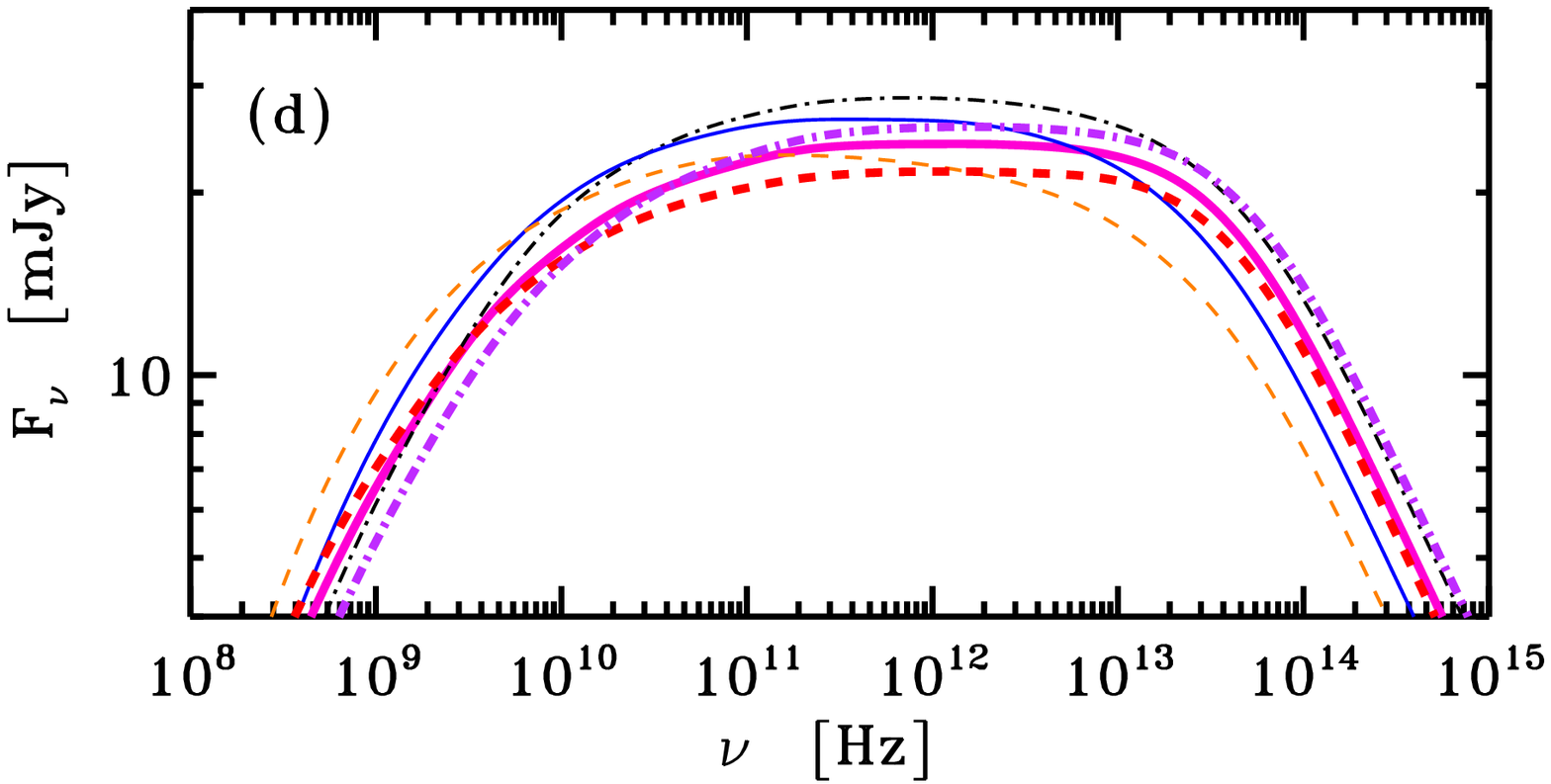}
\caption{Influence of the modelling of shock propagation and dissipation on the resulting dissipation rate (panel a), specific energy (b), volume filling factor (c) profiles and SEDs (d) in the case of a flicker noise fluctuations of the jet Lorentz factor.  The meaning of the different styles of curves is the same as in Fig.~\ref{fig:sin}}
\label{fig:shockprop}    
 \end{figure}

In order to simulate the hierarchical merging of the ejecta and their emission, I use a newly developed numerical code. In this scheme, at each time step $\Delta t$, a new shell of gas is ejected with a Lorentz factor that depends on the time of the ejection.  The injected shells are then followed until they interact and merge with other ejecta. During the propagation, the internal energy of a shell decreases due to adiabatic losses. During the mergers a fraction of the kinetic energy of the shells is converted into internal energy. The product of mergers are in turn followed until they interact again, and so on. The temporal evolution of the initial Lorentz factors of the shells injected at the base of the jet is determined according to its specified PSD shape. This code is similar to that described in JFK10, although developed independently. The reader is referred to JFK10  for a detailed description of the model and method. Here I will focus only on the main differences between my code and that of JFK10.

\subsection{Collisions of ejecta}\label{sec:shellcoll}

Probably the mot important differences regard the treatments of the shell collisions. Let us consider two colliding  shells of masses $m_1$ and $m_2$ travelling with Lorentz factors $\gamma_1$ and $\gamma_2$ respectively, corresponding to velocities $\beta_1c$ and $\beta_2c$. The Lorentz factor of the merged shell correponds to the Lorentz factor of their centre of momentum:
\begin{equation}
\gamma_{c}=\frac{m_{1}\gamma_1+m_{2}\gamma_{2}}{\sqrt{m_1^2+m_2^2+2m_1m_2\gamma_1\gamma_2\left(1-\beta_1\beta_2\right)}}
\end{equation}
This differs from the  formula used by JFK10 to evaluate $\gamma_c$ (their equation 11) which is valid only in the ultra-relativistic limit. 
The amount of kinetic energy that is converted into internal energy during the collision is then:
\begin{equation}
E_{s,i}=(\gamma_1m_1+\gamma_2m_2)c^2-\gamma_c(m_1+m_2) c^2.
\end{equation}

When two shells collide, a forward shock and a reverse shock form at their interface and then propagate in opposite directions in the frame of the centre of momentum of the shells. Their motion is followed in the simulation.
As in JFK10, the velocity of the shock fronts  is calculated using the jumps conditions of Blandford \& McKee (1978).  However, contrary to the model of Blandford and McKee (1978) the shells are in expansion, and in order to approximate the effect on the shock propagation velocity  and also in order to keep the shock crossing time finite, it is assumed that the shock front velocity is the velocity used by JFK10 plus the shell expansion velocity.  During this process the shell is gradually compressed, and its length decreases until the shock has crossed the shell.  In contrast, JFK10 considered instantaneous compression i.e. at the time of contact the length of the shell is changed to the size it will have at the end of the compression. The present treatment is more realistic and the compression is gradual.  

However, it is often the case that  multiple collisions  with other shells occur over the shock crossing time. Therefore a shell may be traversed by many shock fronts traveling in different directions in its rest frame. Some shock fronts may encounter other shock fronts. For simplicity and in order to limit the number of shock fronts that are followed numerically, I assume that when two shock fronts meet, the one formed during the earlier collision disappears. 

Between two events, the instantaneous power dissipated at a shock front is assumed to be a constant $P_s=E_{s}/t_{dis}$. Where $E _{s}$ is the remaining energy available for dissipaion in the shock. And $t_{dis}$ is the dissipation time. 
 $t_{dis}$ is chosen to be the time before the shock front will reach the outer boundary of the ejecta. If  two shocks meet before reaching a boundary, then the energy of the surviving shock front becomes the sum of the respective energies of both shock fronts  reduced by the amount of energy dissipated up to that point:
 \begin{equation}
 E_{s,1,a}=   E_{s,1,b}(1-{\Delta t_1 }/{t_{dis,1}})+E_{s,2,b}(1-{\Delta t_2}/{t_{dis,2}})
 \end{equation}
 where the indices 1 and 2 refer to the parameters of the 'younger' (surviving) and 'older' (disappearing)  shock front respectively, and the index  $b$ and $a$ denote the properties of the shock before and after the encounter, respectively. $\Delta t_1$ and  $\Delta t_2$ are the time since last update of the shock front energies.  $E_{s,1,b}$ and $E_{s,2,b}$ are the energies remaining to be dissipated just after the last update. At the same time, $t_{dis,1}$ is updated according to its  current distance to the shell boundary and the shock front and current boundary velocities.  

 Since the shock fronts that meet each other tend to travel in opposite directions (in the CM frame), transferring the remaining energy of the suppressed shock front to the surviving one leads to overestimate the energy dissipation time.  Indeed some residual energy is trapped and may have to cross the shell several times carried by shocks traveling in different directions. To investigate how this affects the results, I have implemented another scheme in which all the shock energy is dissipated before it either reaches the boundary, or interact with a 'younger' shock front. In this scheme, $t_{dis}$ is the time required to reach the boundary or encounter another shock, whatever is the shortest. As soon  as,  during the shock propagation,  something happens to the shell (e.g. a collision occurs), $t_{dis}$  is updated accordingly and $E_{s}$ is reduced by the amount of energy that was dissipated up to this point. Therefore, in this scheme the energy of a shock is dissipated only during its lifetime. Since, for technical reasons,  we arbitrarily make shocks disappear, this method  tends to underestimate the dissipation time-scale.  For this reason I call this scheme the `fast' dissipation method, and I call `slow' dissipation method the scheme described in the previous paragraph. 
  
During the shock propagation and compression the shells are also expanding radially and loose energy through adiabatic expansion.  The internal energy of a shell with centre of mass initially located at a distance $z$  from the base of the jet with an internal energy $u_{int}$, traveling with a velocity $v$ and which is traversed by shock fronts that dissipate a constant total power $P_{T}$, is after a time  $\Delta t$:
 \begin{equation}
u_{int}(\Delta t)=\left[\frac{P_{T}}{2\gamma_a-1}\frac{z}{v}\left(x^{2\gamma_a-1}-1\right)+u_{int}(0)\right] x^{2-2\gamma_a}
\end{equation}
where $x=1+\frac{v\Delta t}{z}$. Note that the internal energy in the co-moving frame is $\tilde{u}_{int}=u_{int}/\gamma$.
In the simulations, the shock energy is  dissipated only if the relative velocity of the two shells is larger than the local sound speed. This is to account for the fact that only supersonic collisions can lead to shock formation and particle acceleration. However this turns out to have practically no effects on the results because the energy of subsonic collisions is negligible.
 
 \subsection{Shell injection and generation of input Lorentz factor fluctuations}\label{sec:shellinjection}

The standard method to generate time series with the desired PSD is the method of Timmer \& K\"onig (1995). This method was designed to model power-law noise but is easily generalised to any PSD shape.  For a process with a given PSD, this method allows one to generate a realisation of the Fourier transform through a random sampling of the squared amplitude and phase at each discrete Fourier frequency. Then, the time-series is generated by applying the inverse Fourier transform.  The amplitudes are sampled according to a $\chi^2$ distribution with 2 degrees of freedom, while the phases are drawn from a uniform distribution.  This method takes into account the full stochasticity of a linear process. However due to the random sampling of amplitudes, a given realisation may have a power spectrum which differs significantly from that of the parent process. This may affect the results of the simulation in a way that is uncontrolled. In order to determine the average effect of a process with a given power spectrum it is often more convenient to  fix the amplitude of the realised Fourier spectrum to that of the parent process and randomise only the phases (as done e.g. in Done et al. 1992). This is the method that we will use in this paper, although both methods are implemented in the code. 

The time-series generated according to these methods have fluctuations that are symmetric with respect to the average. Since we are often dealing with large amplitudes fluctuations,  the time series can occasionally take negative values.  This is a problem if the time-series is supposed to represent the evolution of the Lorentz factor. If the number of negative values is small one can just replace the negative values by an average of the neighbouring (positive)  values.  If the number of negative values is large, another technique consists in generating a time series $l_i$ of desired PSD as before but with  a mean of 0. Then define the Lorentz factor as:
\begin{equation}
\gamma_i=1+(\gamma-1)\frac{\exp{l_i}}{\langle\exp{l_i}\rangle},
\end{equation}
where $\gamma$ is the desired average value of the Lorentz factor and $\langle\exp{l_i}\rangle$ denote the mean value of $\exp{l_i}$. 

This procedure generates Lorentz factors that never fall below unity. The PSD of  $\gamma_i$  usually remains very close to that of $l_i$.
I also note that taking the exponential of $l_i$ makes the time series non-linear (Uttley, McHardy \& Vaughan 2005). For this reason, in the following,  this method of generating the Lorentz factors will be referred to as the `non-linear  method' while the simple direct method will be called 'the linear method'. In fact the non-linear method is probably more realistic, because non-linearity appears to be a general feature of the variability of accreting black holes (Uttley et al. 2005). For this reason this is the method that is used by default in this paper.  A comparison of the results obtained with both methods will be discussed in Section~\ref{sec:results}.

 In most of the simulations shown in this paper I assume that the shells are all injected with the same mass.  In principle, their mass could also fluctuate. Besides the case of constant mass injection the code also allows one to simulate models in which the shells are injected with a constant kinetic power, and a third case in which the mass of the shell is variable with the same PSD as the Lorentz factor but the mass fluctuations and Lorentz factor fluctuations are generated independently and are uncorrelated. These different scenarii will be discussed in Section~\ref{sec:results}.
 
 \subsection{Radiation}

The time dependent emission of each shell is calculated as described in Appendix~\ref{sec:numem}. This treatment of radiation transfer is very simplified and I summarise here the main approximations.

First, the anisotropy of radiation in the co-moving frame of the ejecta is not taken into account. For example, equation~(\ref{eq:inu}) neglects the angular dependence of the optical depth. This is a good approximation only for systems with a large inclination. Also, the contribution of each individual active ejecta is computed independently, without taking into account the illumination from its neighbours. This effect might be important at small jet inclination when the line of sight crosses many ejecta. 

Moreover, the spatial distribution of particles and magnetic field in the shells are assumed to be homogenous and the electron energy distribution is a pure power-law.  The effects of the thermal component in the electron distribution (see Pe'er \& Casella 2009) are neglected. The effects of adiabatic cooling are taken into account by changing the normalisation of the electron distribution while its effects  on the shape of the electron distribution are neglected. The radiation from the surface of the ejecta is assumed to be uniform at all times, the delays due to photons propagation within a shell are not taken into account.   Simple estimates, indicate that in cases that are relevant to the observations, these effects alter the SED only marginally. 
 
On the other hand, the effects of radiative cooling  can be very strong in the region close to the base of the emitting region (see e.g. Chaty et al. 2011). These radiation losses are neglected in the present version of the model. They affect predominantly the most energetics electrons, emitting in the optically thin regime, while the flat optically thick part of the SED  is essentially unchanged (Zdziarski et al. 2013). 

 Also, only the synchrotron emission is calculated and other processes such as inverse Compton emission are neglected. Inverse Compton scattering of synchrotron photon or external radiation can be a significant source of gamma-ray emission which I have neglected. A comparison of the predictions of the internal shock model with observations from high energy instruments such as INTEGRAL, Fermi or HESS would certainly put additional constraints on the model. 
 
 The treatment of radiation transfer could therefore be improved and this is devoted to future works. 
In fact, better calculations of the emission from two colliding ejecta have been proposed in the context of AGN (see  e.g. in  Joshi \& B\"{o}ttcher 2011; Jamil \& B\"{o}ttcher 2012).  In the case of X-ray binaries however, due to the much shorter time scales,  it is necessary to add up the contribution of millions of shells in order to be able to compare with the observations. Such a detailed treatment of radiation transfer in the compact jets of X-ray binaries would be therefore extremely time consuming, while unlikely to change qualitatively the results.

Finally it is worth noting that other possible dissipation mechanisms such as magnetic dissipation or other types of shocks could be important. In particular, in the case of high mass X-ray binaries such as Cygnus X-1 or Cygnus X-3, the interaction of the jet with the wind of the companion star may lead to the formation of re-collimation shocks (Perucho, Bosch-Ramon \& Khangulyan 2010). Such additional dissipation would obviously add some spectral complexity.

\begin{figure}
\includegraphics[width=\linewidth]{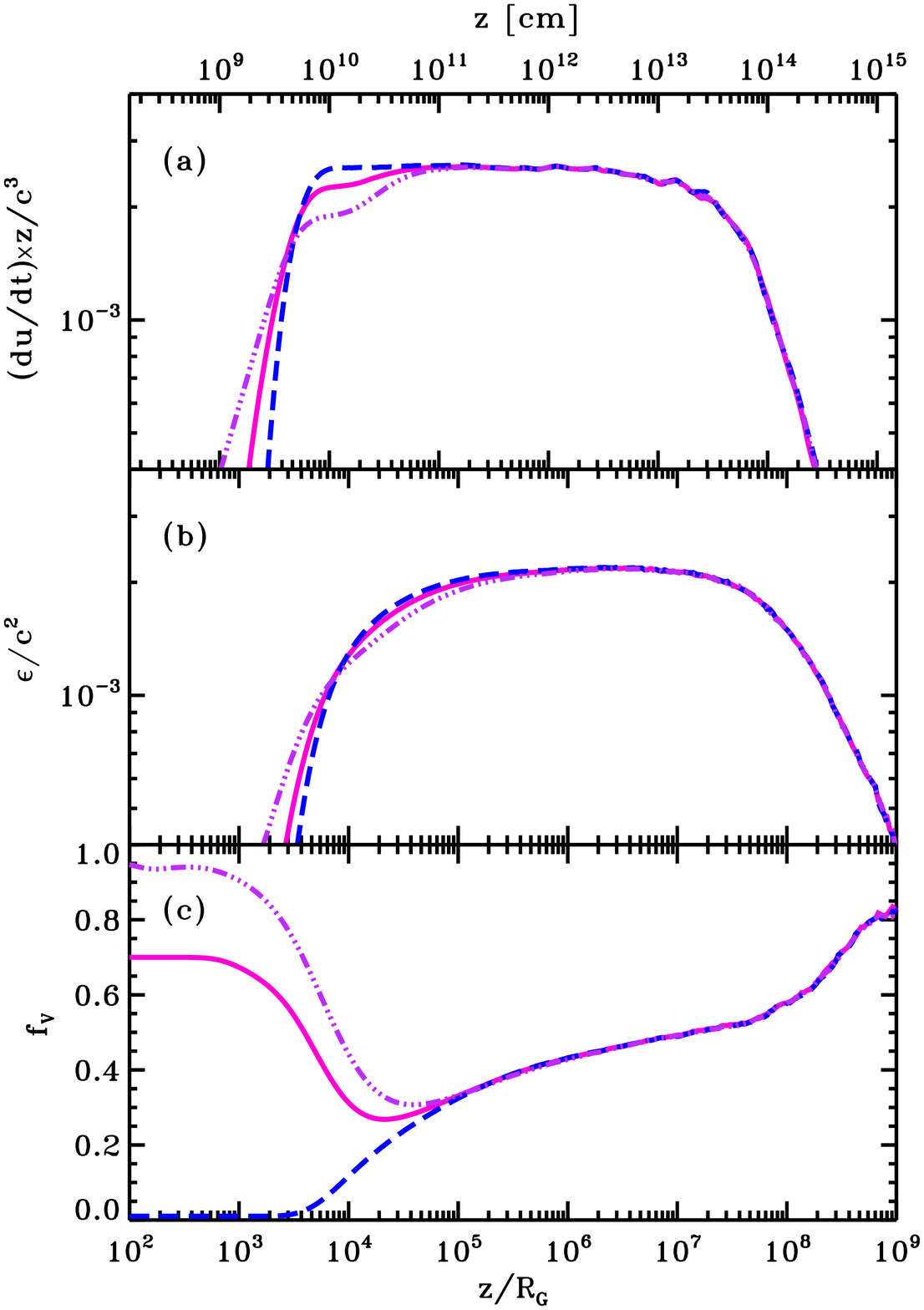}
\includegraphics[width=\linewidth]{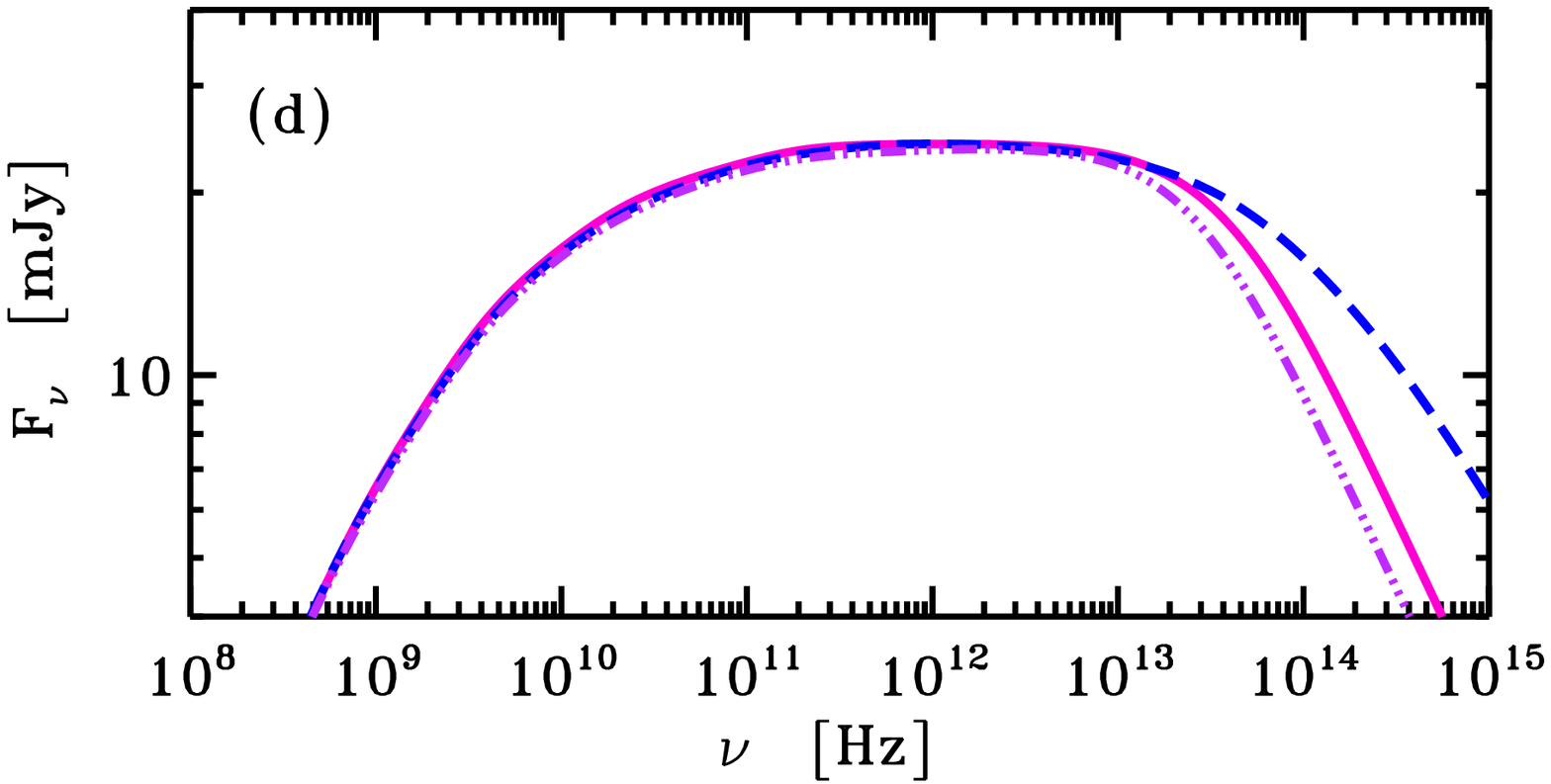}
\caption{Influence of the initial volume filling factor of the shells at injection on the resulting dissipation (a), specific energy (b), volume filling factor (c) profiles and SEDs in the case of a flicker noise fluctuations of the jet Lorentz factor
 The full curves show the result for the baseline model (with initial volume filling factor $f_{\rm v0}=0.7$) the long dashed curves show the results for $f_{\rm v0}=10^{-2}$ and the 3-dots-dash curves for $f_{\rm v0}=0.95$.
 The other parameters are identical to that of Fig.~\ref{fig:shockprop}.}\label{fig:varfv} 
  \end{figure}

\section{Results}\label{sec:results}

{In this section, I explore the model for jet parameters that are typical of X-ray binaries. I consider a stellar mass black hole of 10${M}_{\sun}$.  The average jet Lorentz factor is $\gamma=2$ for a kinetic power $P_J=10^{-2} L_E$ (where $L_E=1.3 \times 10^{39}$ erg/s  is the Eddington luminosity). The jet  half opening angle is $\phi=1\degr$.  The equipartition ratios are $\xi_e=1$, $\xi_{\rm p}=0$, the electrons in the jet have a power-law energy distribution with slope $p=2.3$ between $\gamma_{\rm min}=1$ and  $\gamma_{\rm max}=10^{6}$. The jet radius at the base is $R_b=10$ gravitational radii $(R_G)$. }

\subsection{Response to sinusoidal variations of the Lorentz factor}\label{sec:sin}

As a first example I study here the simple case where the Lorentz factor varies with time like a sine function. 
The sine oscillation is probably irrelevant for the observations but is nevertheless interesting  to understand the response of the jet to fluctuations of a well defined time scale $T_i=1/f_i$ (or equivalently lengthscale $\lambda=\beta cT_i$) and amplitude of velocity fluctuations $\Delta v$.  In this case, most of the dissipation is expected to occurs around one typical distance $z_d  \propto \lambda/\Delta v$.
Longer  time-scale (i.e. smaller $f_i$) fluctuations have a larger spatial extension $\lambda$  i.e. ejecta of significantly different velocities are ejected with a greater separation. A longer separation implies that it takes more time for the ejecta to catch-up, leading to dissipation at larger distance in the jet. On the other hand, fluctuations of larger amplitude imply a larger velocity difference of the colliding shells and therefore collision at a shorter distance. 

Using the formalism developped in Section~\ref{sec:analytic}, I can now estimate  $z_d$, as well as the specific energy and magnetic field profiles. 
Let us assume sinusoidal fluctuations of the Lorentz factor at the frequency $f_i$ so that the PSD of the fluctuations is given by $S(f)=\gamma_{\rm rms}^2 \,\delta(f-f_i)/2$, where $\delta$ is the Dirac function.  Then equation~(\ref{eq:dusdk}) gives the energy dissipation as:
\begin{equation}
\frac{d\tilde{u}}{dK}=-\frac{f_0\gamma_{\rm rms}^2 \delta(f_0/K-f_i) c^2}{2 K^2\gamma^2\beta^2}.
\label{eq:dusdksin}
\end{equation}
From equations~(\ref{eq:dv2}) and~(\ref{eq:jk}),  the average velocity of the collisions is:
\begin{equation}
\Delta\tilde{v}=\frac{\gamma_{\rm rms}c}{\gamma\beta }\sqrt{2 \left[1-\cos{(K \pi f_i/f_0)}\right]},
\label{eq:dvsin}
\end{equation}
for $K \le f_0/f_i$,  and $\Delta\tilde{v}=0$ otherwise. 
Equations~(\ref{eq:tdek}),~(\ref{eq:dusdksin}) and~(\ref{eq:dvsin}) imply that all the energy of the fluctuation is dissipated at a time:
\begin{equation}
\tilde{t_{d}}=\tilde{t}(f_0/f_i)=\frac{y\gamma^2\beta^2}{4\gamma_{\rm rms}f_i},
\end{equation}
This corresponds to a distance from the base of the jet:
\begin{equation}
z_d=\frac{y\gamma^3\beta^3c}{4\gamma_{\rm rms}f_i},
\label{eq:zd}
\end{equation}
and, under the assumption of a conical jet, this corresponds to a jet radius:
\begin{equation}
R_d=R_b+z_d \tan{\phi} 
\end{equation}
After dissipation, the specific internal energy decreases according to adiabatic losses. For negligible initial internal energy ($\tilde{\epsilon}_0\simeq0$), Equation~(\ref{eq:evole}) gives:
\begin{equation}
\tilde{\epsilon}=\frac{\gamma^2_{\rm rms} c^2}{2\gamma^2\beta^2}\left(\frac{R}{R_d}\right)^{-2(\gamma_a-1)},
\label{eq:epsin}
\end{equation}
for $R\ge R_d$, and  $\tilde{\epsilon}=0$ otherwise.
And therefore  at  $R\ge R_d$ the magnetic field decays like:
\begin{equation}
B=\left[\frac{2 P_J}{(1+\xi_{e}+\xi_{\rm p})f_{\rm v}(\gamma-1) \gamma^3\beta^3c}\right]^{1/2}\frac{\gamma_{rms}}{R_d}\left[\frac{R}{R_d}\right]^{-\gamma_a}.
\label{eq:bsin}
\end{equation}
From equations~(\ref{eq:dvsin}) and (\ref{eq:epsin}), the Mach number of the collisions around $z_d$ is:
\begin{equation}
\mathcal{M}=\frac{\Delta{\tilde{v}}}{\sqrt{(\gamma_a-1)\tilde{ \epsilon}}}=\frac{2\sqrt{2}}{\sqrt{\gamma_a-1}}\simeq\sqrt{24}.
\end{equation}
Therefore the collisions at $z_d$ are supersonic ( $\mathcal{M}\simeq\sqrt{24}$) and shock acceleration may indeed occur. 
According to equations~(\ref{eq:fe}) and ~(\ref{eq:y}),  the volume filling factor is set to  $f_{\rm v}=f_e\simeq 0.26$ and $y\simeq 0.53$.  

 An example is shown in Fig.~\ref{fig:sin} for sinusoidal variations of the Lorentz factor at a frequency  $f_i=1$~Hz and an amplitude $\gamma_{\rm rms}=0.3$.  The shells are ejected at time intervals of 10 ms with an initial volume filling factor $f_{\rm v0}=0.7$ and no internal energy.  
Panels (a), (b) and  (c)  of Fig.~\ref{fig:sin}  display the power dissipated through internal shocks, the specific energy of the flow and the volume filling factor of the ejecta respectively, as a function of the distance from the point of ejection. 
 Equation~(\ref{eq:zd}) gives $z_d\simeq 5\times 10^{4}\quad R_G$,  as shown by the vertical dotted line in Fig.~\ref{fig:sin}(a). 
At $z_d$,  the internal specific energy increases sharply. Then, at larger distances, it decays like $z^{-2/3}$, due to adiabatic losses. This is  shown by the thick dotted line in panel (b) of Fig.~\ref{fig:sin}.  From the magnetic profile given by  equation~(\ref{eq:bsin}), the analytical estimate of the SED is derived as described in Appendix~\ref{sec:plbprof} assuming a constant $f_{\rm v}\simeq0.26$. It is shown in Fig~\ref{fig:sin}(d). At low frequencies the partially self-absorbed component has a predicted spectral slope $\alpha_{\rm T}=0.65$ , the spectral break occurs around  $5\times 10^{12}$ Hz. 

 Figure~\ref{fig:sin} also compares the above analytical estimates with the results of simulations using several different prescriptions for the internal dynamics of the shells and the associated dissipation processes. 
The profiles shown in Fig.~\ref{fig:sin} were estimated by averaging the properties of the ejecta at a given location, over a time-span ranging from 10$^3$ to 10$^4$s after the ejection of the first shell.  The  time-averaged SED of the jet was  obtained by averaging the synchrotron emission of the ejecta over the same time interval.

The thin dot-dashed lines present the results for the simplified case in which the shells behave as solid blocks which cannot expand or compress ($f_{\rm v}$ is constant), and all the shock dissipation occurs instantly at the time of contact between two colliding shells.  In this case the shock dissipation profile is sharp and peaks around 2$\times 10^4$ $R_G$.
This leads to a sharp rise in the specific energy of the flow around this distance. The specific energy profile then decays as expected in the analytical model.  The resulting SED has the analytically predicted  slope at low frequency and peaks around $10^{13}$ Hz.

The thick dot-dashed curves show the result of the same simulation but now including the effects of the compression and expansion of the ejecta. Fig~\ref{fig:sin}(c) shows that at short distances from the black hole, the volume filling factor remain constant due to the absence of compression or internal pressure. Then around $10^4$ $R_G$, the first shocks occur and  compress the ejecta ($f_{\rm v}$ decreases down to $\simeq 0.4$). This compression delays further collisions and the dissipation profile (panel (a)) is broader than before and peaks at larger distances. After the bulk of the dissipation, the ejecta start to expand under the effects of their own pressure. They then continue to  expand until they fill the whole volume.  As a consequence of the delayed dissipation the specific energy profile rise is smoother than in the previous cases and peaks at  about 3 times larger distances, before decaying as $z^{-2/3}$. The delayed dissipation leads to a SED  that is  slightly shifted toward lower frequencies although quite similar to that obtained in the previous case.

In more realistic simulations the energisation of the ejecta is further delayed due to the spread of the dissipation over the shell merging time (or shock propagation time). The full thick curve shows the result obtained in this case and assuming the `fast' dissipation prescription (see Section~\ref{sec:shellcoll}). The dissipation profile is now somewhat broader and shifted toward larger distance. But overall, the specific energy profile and SED remain very similar to what obtained when the dissipation time was neglected. 
On the other hand, the `slow' dissipation prescription (shown in thick dashed curves)  leads to a much broader dissipation profile. It is  formed by a first bump peaking around $10^5$ $R_G$ plus an extension that reaches the maximum distance of a few $10^8$ $R_G$ attained by the shells during the simulation time. This extension  is due to the constant dissipation rate of the shock energy that remains trapped in the shells as a consequence of the `slow' dissipation prescription.  As a consequence, the peak of the specific energy occurs at a longer distance and at a level which is a few times lower than in the `fast' dissipation case. As the internal energy is lower, the compression is stronger. The volume filling factor reaches lower values and does not recover as quickly as before after compression.      
Also, due to the presence of the residual energy dissipation at large distances the specific energy does not decay as $z^{-2/3}$  but is flat or even slightly rising after $10^7$ $R_G$.
As shown in panel (c),  this causes the SED to flatten at long wavelengths. The sharp drop  in flux below 1 GHz is related to the distance attained by the first ejecta in the course of the simulation time.

Interestingly, when the dissipation time is taken into account but the shells are not allowed to compress and expand, the dissipation is even more delayed (as shown by the thin full and dashed curves in Fig.~\ref{fig:sin}). This is because, in these cases,  the absence of compression makes the dissipation time-scale longer. 

Regarding the most realistic simulations taking into account both the internal dynamics of the ejecta and the dissipation time scale, the result seem to depend critically on the type of prescription for the dissipation time scale. The `fast' dissipation assumption, which probably underestimates the dissipation time-scale, produces dissipation over a relatively narrow range of distances and the results are in qualitative agreement with the predictions of the analytical model.  On the other hand, the `slow' dissipation prescription, which overestimates the shock dissipation time-scale, spreads the dissipation over  the whole jet extension. This affects significantly the shape of the predicted SED.  However we will 
show in the next section that the differences obtained under `fast' and `slow' dissipation assumptions are strongly reduced if, instead of a simple sinusoidal variation of the jet Lorentz factor, one considers a mixture of fluctuations with widely different time-scales and amplitudes.

\begin{figure}
\includegraphics[width=\linewidth]{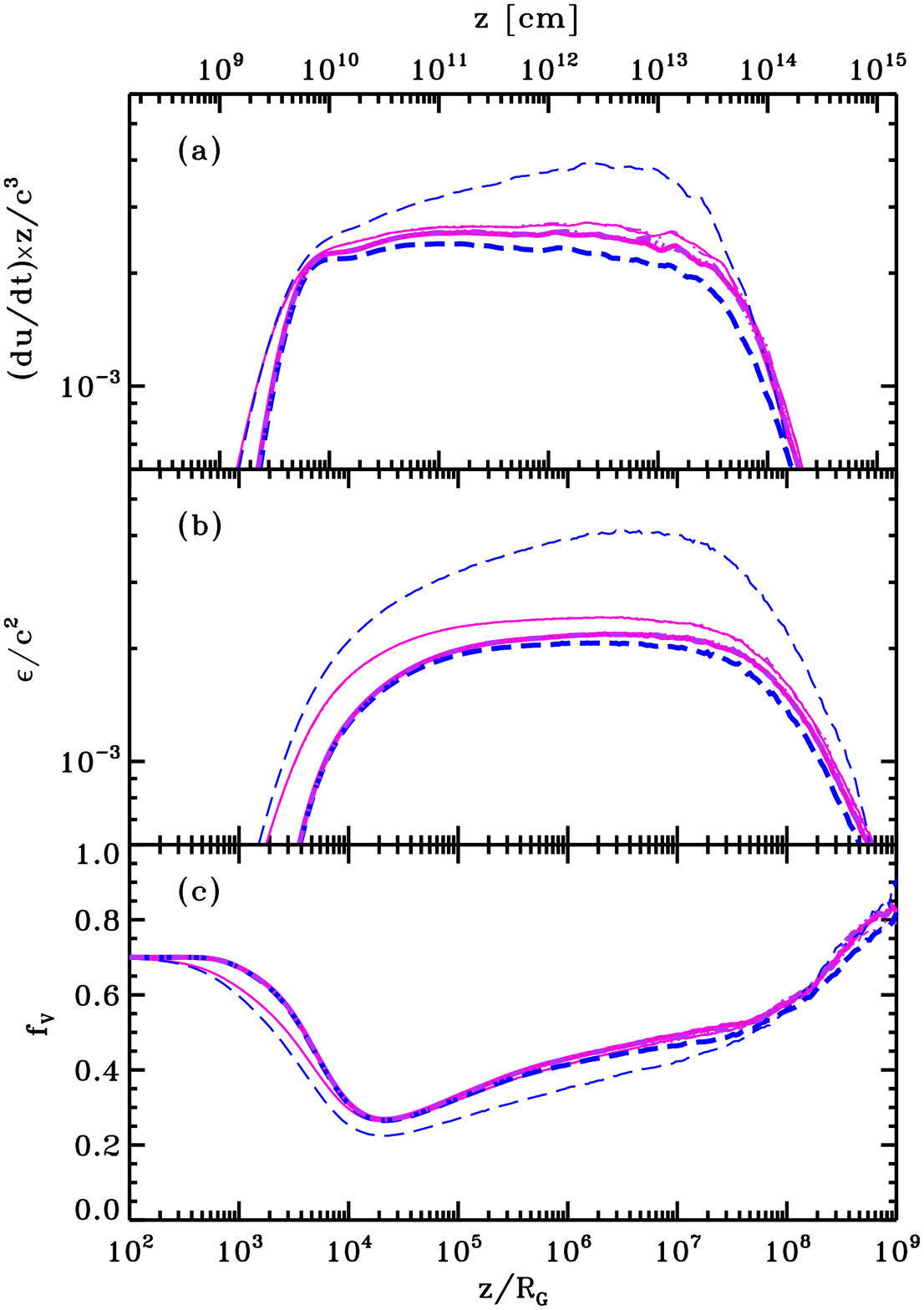}
\includegraphics[width=\linewidth]{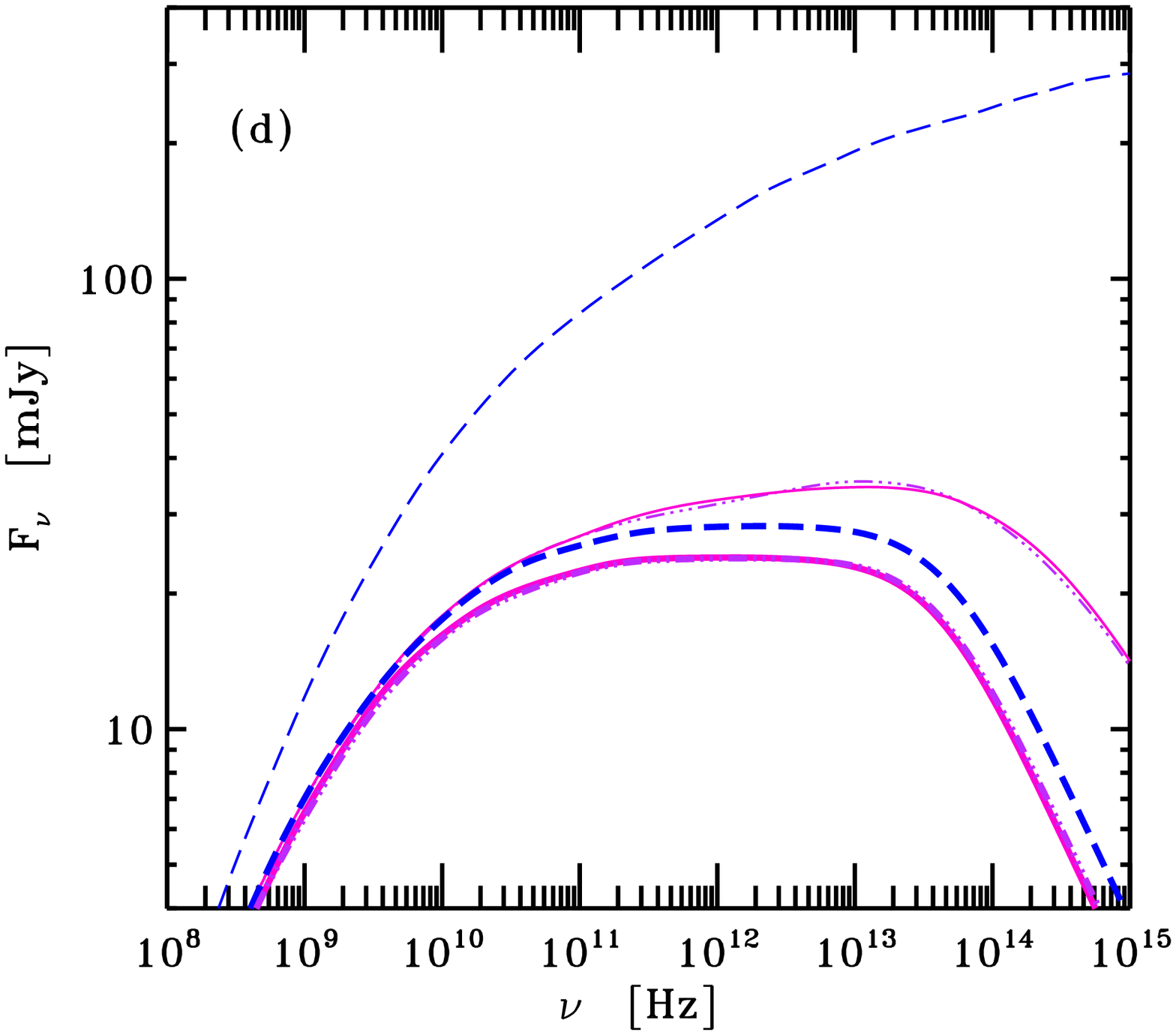}
\caption{Influence of the modelling of the input fluctuations. 
 The thick and thiner curves show the result of simulations obtained with the `non-linear' and 'linear' input fluctuations respectively (see text). The full curves show the results for a constant mass of the ejecta. The  triple-dot-dash curves, almost undistinguishable from the full curves, are obtained for a  mass of the ejecta that is varying randomly and independently of the Lorentz factor. The long dash curves show the results obtained  for a constant kinetic energy of the ejecta. The other parameters are identical to that of Fig.~\ref{fig:shockprop}}\label{fig:linnonlin}
\end{figure}

\begin{figure*}
\includegraphics[width=\linewidth]{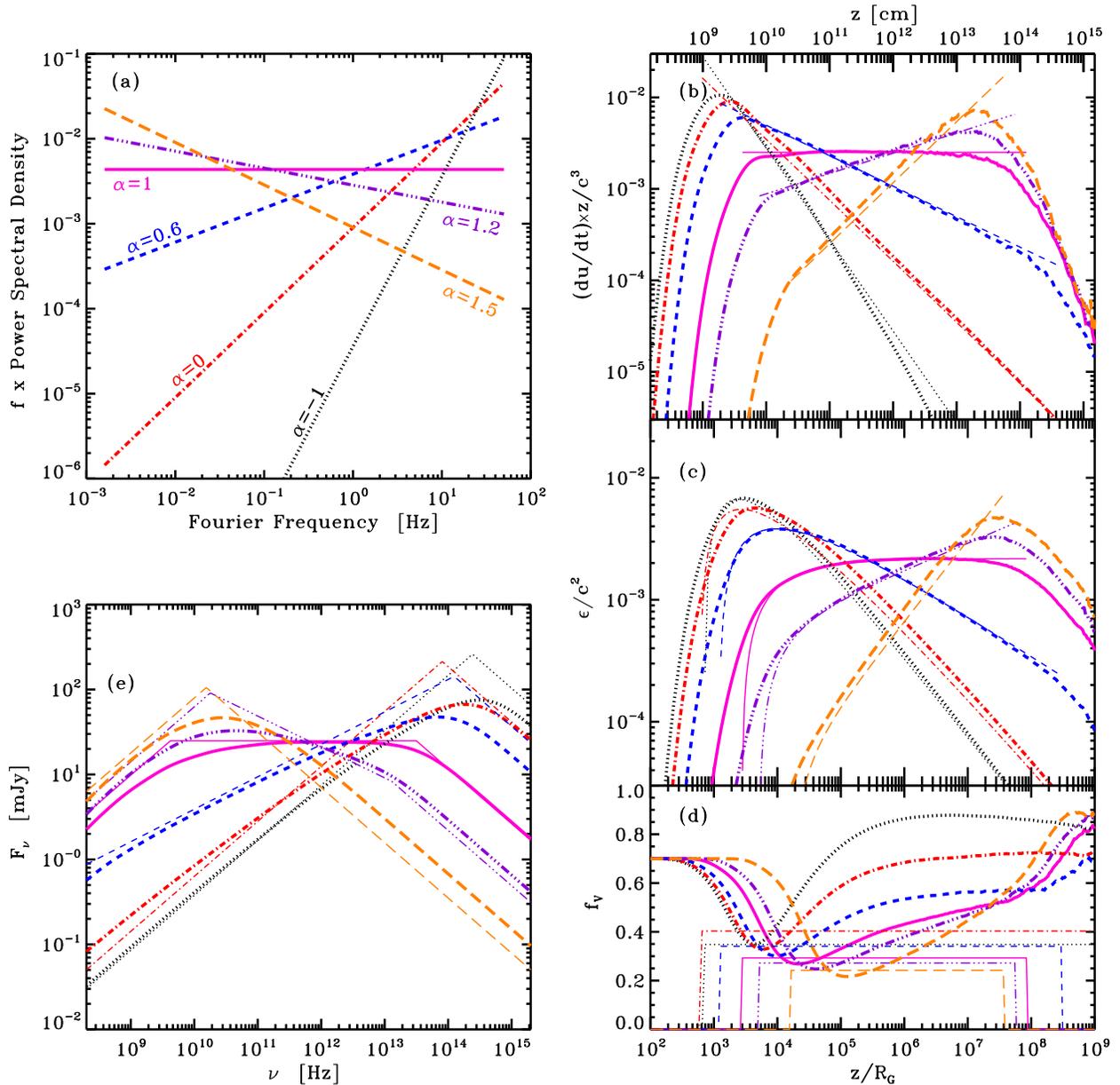}
\caption{Simulations of the internal shock model with a power-law PSD of the Lorentz factor fluctuations ($P(f)\propto f^{-\alpha}$).  Panel (a) shows the shape of the injected PSDs, for the indicated values of the $\alpha$ index. Then moving clockwise other panels show the time averaged dissipation  (b), the specific energy (c) and volume filling factor (d) profiles.  The SEDs are shown in panel (e). The thin curves show the analytical estimates.  } \label{fig:1} 
\end{figure*}

\subsection{Flicker noise fluctuations}\label{sec:flickernoise}

In this section, I study the case of  fluctuations of the jet Lorentz factor which have  a flicker noise PSD (i.e. $P(f) \propto 1/f$) over a wide range of Fourier frequencies. This case was discussed extensively in M13 from the point of view of the comparison of analytical estimates to observations. Here, I investigate further the effects of the different options for the numerical modelling of the shock propagation and injected fluctuations that were presented in Section~\ref{sec:numeric}.  The baseline model is the same as that presented in M13. 
Fig.~\ref{fig:shockprop} shows the resulting dissipation, specific energy and volume filing factor profiles, averaged between  $10^3$ and $10^5$ s after the time of the first ejection. 
This figure confirm the analytical results of M13 i.e.  independently of the detail of the model, the energy dissipation profile scales like $z^{-1}$ over a wide range of distances from the black hole, and the specific energy profile is nearly a constant.  As in the simulation with sinusoidal fluctuation the volume filling factor of the shells drops quickly during the first shocks, however the extended range of time-scales of the flicker noise fluctuations lead to persistent dissipation up to large distances. The associated compression prevents the shells from expanding and filling the entire volume until all dissipation ceases. This occurs only when the longest fluctuations have been dissipated and at about $10^8$ $R_G$, see panel (a). Closer to the black hole,  the volume filling factor varies very weakly with $z$ and remains in the range 0.3-0.6 over about 4 orders of magnitude in distances.  
The resulting synthetic SED is shown in Fig.~\ref{fig:shockprop}(d).  As discussed in M13 the SED is approximately flat at frequencies ranging from a few $10^{9}$ to $10^{13}$ Hz. In the mid-IR the model predicts a spectral break. At higher frequencies the spectral index of the spectrum becomes $(p-1)/2$  as the jet emission is dominated by the optically thin emission from the base of the dissipation region. After travelling a large distance from the black hole there is a point where all the free energy of the ejecta has already been dissipated and no shock dissipation can power the jet emission. The electrons just cool down through adiabatic expansion. This is the cause for the low photon frequency break around a few GHz below which the jet emission decreases sharply. The location of this break is determined by the lowest Fourier frequency of the injected Lorentz factor fluctuations (see M13).

Fig. ~\ref{fig:shockprop} compares the results obtained assuming the `fast', `slow'  and 'instantaneous' dissipation prescriptions (see Sections~\ref{sec:shellcoll} and~
\ref{sec:sin}). The differences between the three cases are modest (10 percent at most). 
Fig.~\ref{fig:shockprop}, also presents the results of simulations in which the ejecta do not expand or compress longitudinally  but have a fixed length which was fixed assuming the equilibrium volume filling factor given by equation~(\ref{eq:fe}). Again, the effect on the shape of the SED is weak. The result is not very sensitive to the treatment of shock propagation and the analytical estimate of $f_{\rm v}$ is reasonably accurate for the purpose of calculating the SEDs. 
The conclusion is that although some assumptions and approximations are required to model the treatment of the shock propagation and energy dissipation in the ejecta and the associated compression, in the case of flicker noise, the resulting SEDs appears to be rather insensitive to the details of this modelling. In the following we adopt the `fast' dissipation method to explore further the effects of other parameters.  

Fig.~\ref{fig:varfv} shows the effects of assuming different volume filling factor of the ejected shells. It shows that different choices for $f_{\rm v0}$ induce differences in flux of a factor of 2 or more at the  higher (optically thin) frequencies. The synchrotron emission is sensitive to the energy density (in the form of particles and magnetic field) in the ejecta. Since the amplitude of the Lorentz factor fluctuations and the mass of the shells are the same in the different simulations shown in Fig.~\ref{fig:varfv}, the energy dissipated in the shocks is the same. However, in the simulations with lower $f_{\rm v0}$ the first shocks occurs in a smaller volume. Also, due to the smaller scale of the ejecta the energy is dissipated faster and the specific energy rises faster (see panels a and b). The larger energy density then implies a higher synchrotron flux. Nevertheless, after the first shock, the shells are compressed and can also expand. Very quickly the evolution of  $f_{\rm v}$ becomes independent of the initial conditions (see panel c). The optically thick emissions of the different models shown in Fig.~\ref{fig:varfv}(d) are almost identical. Therefore, the choice of $f_{\rm v0}$ can affect the shape of the high frequency turn-over and the optically thin emission, but has no effects at lower frequencies. 

Let us now consider the effects of different prescriptions for the properties of the injected fluctuations. 
Fig.~\ref{fig:linnonlin} shows the results for the baseline model assuming both linear and non-linear Lorentz factor fluctuations. There are significant differences because, for  the moderate jet mean Lorentz factor and the relatively large amplitude of the fluctuations that I assume, the 'linear' method implies that many shells have a very small velocity (as small as zero).  In the non-linear method  such low values are not allowed and the velocity is almost always mildly relativistic at least. The length of an injected shell is $l=vf_{\rm v0}\Delta t$. Mildly relativistic shells will have a length $\sim cf_{\rm v0}\Delta_t$, but  the slow ejecta of the linear model can have a much smaller extension. Those slow shells are quickly hit by other ejecta. As can be seen in panel (a) and (b) the energisation of the shells start earlier in the linear model and is more efficient, dissipating a comparatively large amount of energy in a small volume. The earlier shocks also compress the shells more efficiently (see panel c). The resulting synchrotron emission (shown in panel d)  is therefore significantly stronger in the linear model and particularly in the optically thin regime. Interestingly the analytical estimates are in better agreement with the non-linear model (see M13 for a comparison). This is because the analytical model  does not take into account  the dependence of the length of the shell on velocity (the length of each shell is implicitly fixed at the average value). 

 Fig.~\ref{fig:linnonlin} also shows the effects of having randomly variable fluctuations of the mass of the ejecta. It shows that if the mass fluctuations are uncorrelated to the Lorentz factor fluctuation, this has negligible effects on the shape of the SED, both in the linear and non-linear model. Indeed, in this case the mass fluctuations do not change the dissipation profile in the jet. They just add some variability on the amount of energy dissipated during shocks, but this is averaged out in the time-averaged SEDs. The situation can however be very different if the fluctuations of the mass are in some way related to those of the Lorentz factor. In Fig.~\ref{fig:linnonlin}, I considered the case of both variable Lorentz factor and mass but keeping the kinetic energy of the shells a constant equal to $E_c=\langle\gamma-1\rangle m_0$.  Where $m_0$ is the 'constant' mass of the shell in the fiducial model. Introducing this anti-correlation between mass and Lorentz factor makes the average mass of the shell larger than $m_0$ by $\simeq10 \%$ in the non-linear model and by 20\% in the linear model. Therefore the shock dissipation is enhanced and the jet is brighter. In the case of the non-linear model this enhancement is weak (20 percent at most). But in the case of the linear model there is the additional effect  of the presence of the very slow ejecta.  In the simulation with constant jet power, the slow ejecta not only have a very small size but also a huge mass. Shock dissipation leads to a huge energy density. The effect on the SED is dramatic and the jet is much brighter, the flux can increase by approximately one order of magnitude (see panel d). 
In the following we will consider only non-linear fluctuations as they are more representative of the variability observed in accreting sources. In this case the prescription on the variability of the mass of the shell has only small effects on the resulting SED.

 \begin{figure*}
 \includegraphics[width=8cm]{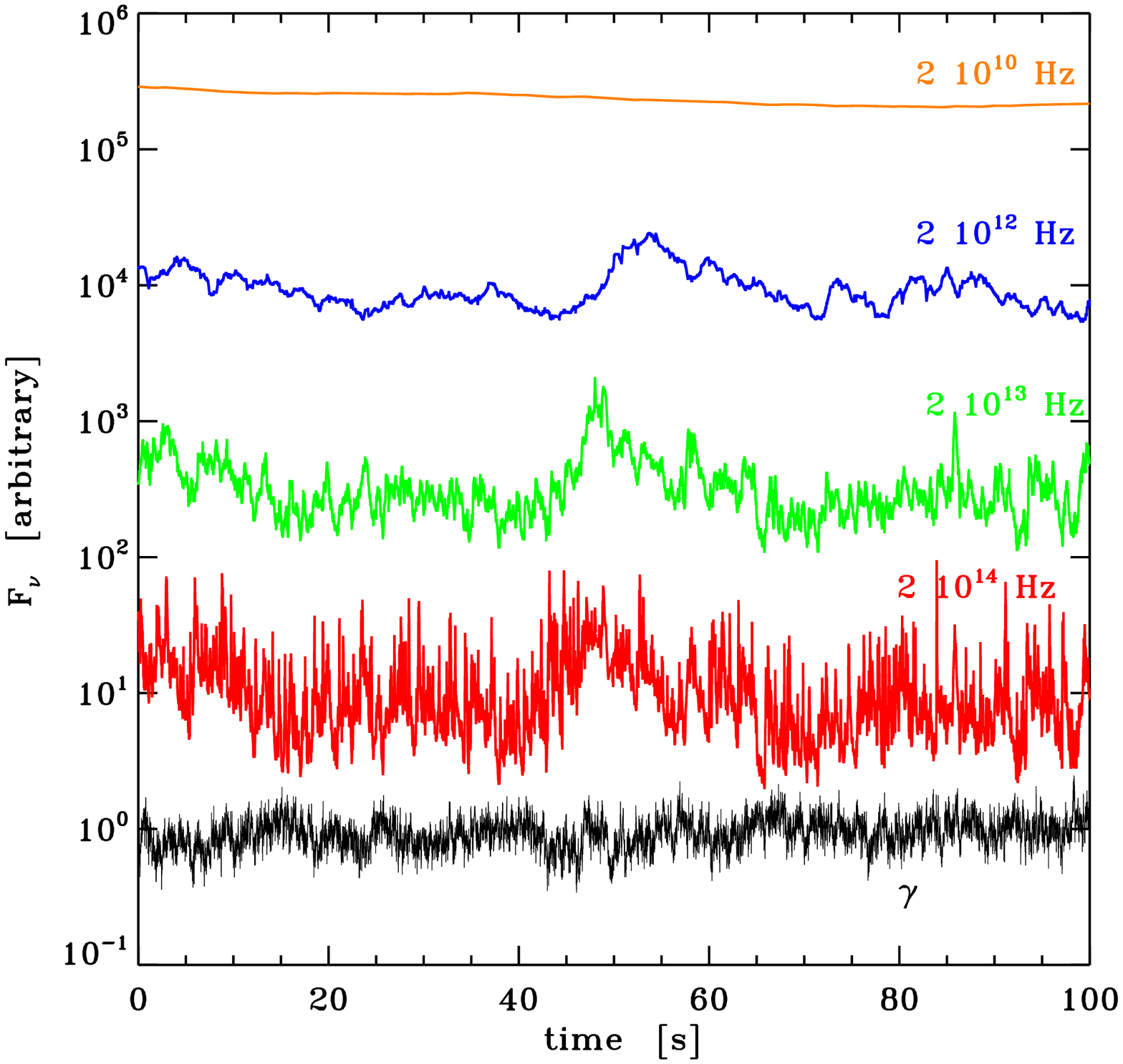}
 \includegraphics[width=8cm]{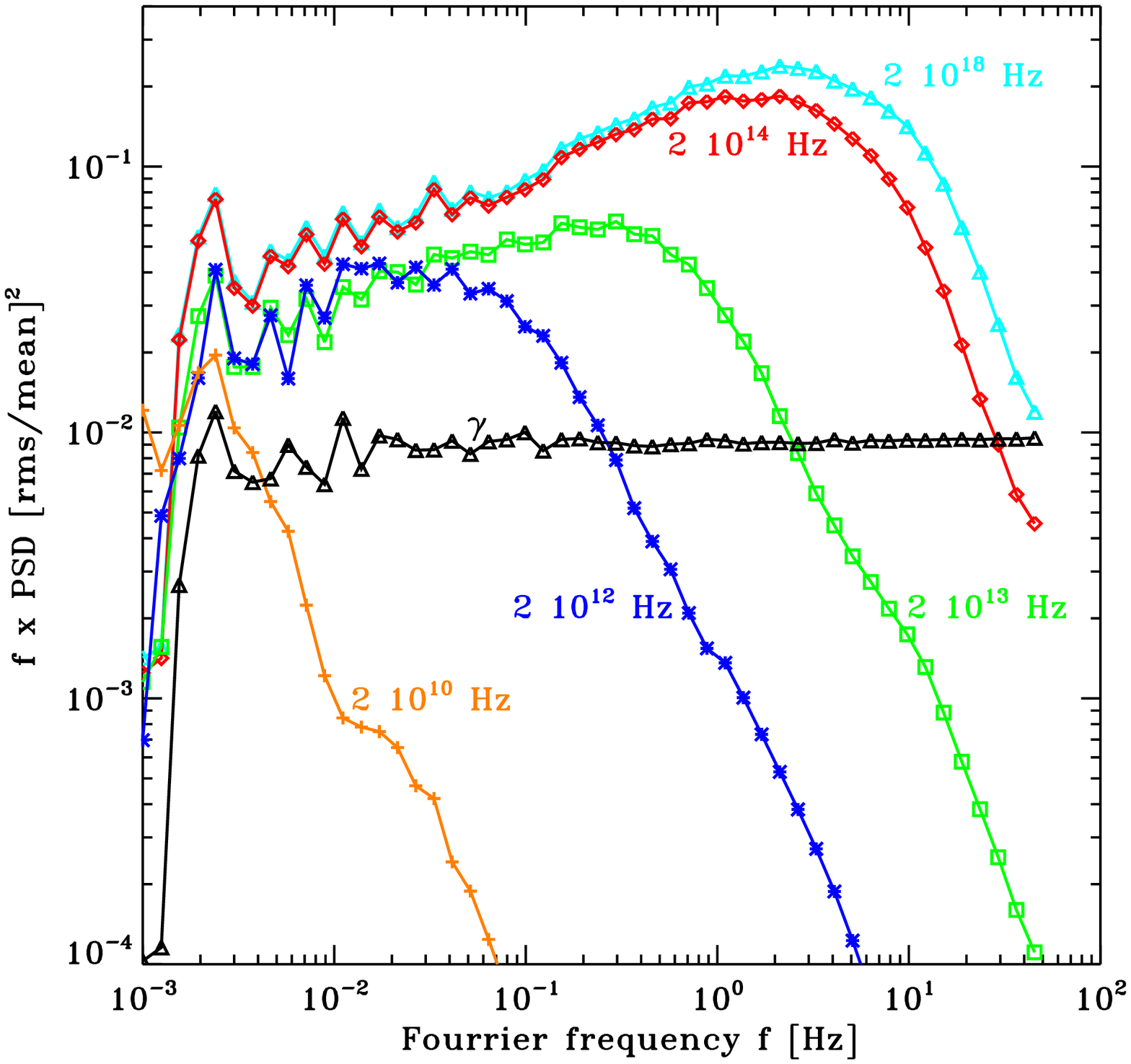}
 \caption{Synthetic light curves (left, rescaled) and power spectra at various  photon frequencies  resulting from the simulation with flicker noise Lorentz factor fluctuations ($\alpha=1$) and a constant jet power. The injected fluctuations of the jet Lorentz factor $\gamma$ are also shown.}
 \label{fig:2}
 \end{figure*}
\subsection{Power-law PSD of fluctuations}\label{sec:powerlawpsd}

Let us now consider Lorentz factor fluctuations with a power-law PSD shape with index $\alpha$: $S=S_0 f^{-\alpha}$  for $f_0>f>f_1$. The other parameters of the model are kept at the same value as in Section~\ref{sec:flickernoise}.  Fig~\ref{fig:1} shows that the dissipation profile along the jet and the profile of the specific energy of the flow are very sensitive to $\alpha$.  For larger $\alpha$ the fluctuations of the Lorentz factor have, on average, longer time-scales  and therefore more dissipation occurs at larger distances from the black hole. As a result the SED is steeper. In the following, the quantitative dependence of the SED on $\alpha$ is derived analytically and compared with the results of the simulations.

\subsubsection{Internal energy and magnetic field profile}

For $\alpha < 3$ the integral $J$ in equation~(\ref{eq:jk}), can be approximated as: 
\begin{equation}
J(K)=S_0G(\alpha)\left(\frac{f_0}{K}\right)^{1-\alpha},
\end{equation}
with:
\begin{equation}
G(\alpha)=-\sum_{n=1}^{+\infty}\frac{(-\pi^2)^n}{(2n)! (2n+1-\alpha)}.
\end{equation}
Then  equation~(\ref{eq:tdek}) gives:
\begin{equation}
K=(\tilde{t}/\tilde{t_0})^q=(z/z_0)^q,
\label{eq:ktz}
\end{equation}
with  $q=2/(3-\alpha)$ and
\begin{equation} 
\tilde{t_0}=\frac{y\gamma^2\beta^2}{4f_0^{1/q}S_0^{1/2}G^{1/2}},
\end{equation}
and $z_0=\gamma\beta c \tilde{t}_0$. 
Then combining equations (\ref{eq:dusdk}) and (\ref{eq:ktz}) and taking $y$ as a constant, we can determine the dissipation profile along the jet:
\begin{equation}
\frac{d\tilde{u}}{d\tilde{t}}=-\frac{q S_0 f_0^{1-\alpha} c^2}{\gamma^2\beta^2 \tilde{t_0} }\left(\frac{z}{z_0}\right)^{\frac{3\alpha-5}{3-\alpha}}.
\label{eq:dusdz}
\end{equation}
This dissipation occurs along the jet until  all the fluctuations have been dissipated, i.e. up to a distance $z_f\simeq z_0 \left(f_0/f_1 \right)^{1/q}$.
Equation~(\ref{eq:dusdz}) is compared to the simulations in Fig.~\ref{fig:1}c and appears to be a good approximation in the range of distances $z_0 < z <z_f$.

In the limit $\tan{\phi}>>R_b/z_0$,  equation~(\ref{eq:evole}) gives:
 \begin{equation}
 \tilde{\epsilon}=\frac{S_0c^2}{\gamma^2\beta^2}f_0^{1-\alpha}K^{(\gamma_a-1)(\alpha-3)} \int_{1}^{K}  x^{(\gamma_a-1)(3-\alpha)+\alpha-2}dx.
 \end{equation}
 For $\alpha=\alpha_c=(3\gamma_a-4)/(\gamma_a-2)$:
\begin{equation}
\tilde{\epsilon}=\frac{S_0c^2}{\gamma^2\beta^2} \left(\frac{f_0}{K}\right)^{-2\frac{\gamma_a-1}{\gamma_a-2}}\ln{K}.
\label{eq:eps1}
\end{equation}
 Otherwise:
 \begin{equation}
 \tilde{\epsilon}=\frac{S_0c^2}{\gamma^2\beta^2} \frac{f_0^{1-\alpha}}{g_\alpha} \left(K^{\alpha-1}-K^{(\gamma_a-1)(\alpha-3)}\right),
 \label{eq:eps2}
 \end{equation}
 and
 \begin{equation}
 g_\alpha=(\gamma_a-1)(3-\alpha)+\alpha-1=(\alpha-\alpha_c)(2-\gamma_a).
 \end{equation}
 
 From this we can estimate the Mach number of the collisions:
 \begin{equation}
\mathcal{M}=\mathcal{M}_0 \left|1-K^{-g_\alpha}\right|^{-1/2},
\end{equation}
where
\begin{equation}
\mathcal{M}_0=\sqrt{\frac{4|g_\alpha|G}{\gamma_a-1}}.
\end{equation}

Equations~(\ref{eq:eps1}) and~(\ref{eq:eps2}) can then be combined to equation~(\ref{eq:ktz}) in order to determine the specific energy profile in the jet. As can be seen on Fig.~\ref{fig:1}d, the result is in excellent agreement with the simulations.
Those equations could also be used to determine the magnetic field profile which in turn can be used to determine the jet SED by numerical integration of equation~(\ref{eq:fnufi}). Here instead, I will provide simple analytical estimates of the SED at the cost of some additional approximations. Indeed, at large $K$ and for $\alpha\neq \alpha_c$ the dominant term  in equation~(\ref{eq:eps2}) has a power-law dependence on $K$:
\begin{equation}
 \tilde{\epsilon}\simeq\frac{ S_0 c^2 f_0^{1-\alpha}}{|g_\alpha|\gamma^2\beta^2}K^{s}.
 \end{equation} 
where $s=\alpha-1$  if $\alpha>\alpha_c$, and 
 $s=(\gamma_a-1)(\alpha-3)$  if $\alpha<\alpha_c$.
In this approximation,   the Mach number is a constant ($\mathcal{M}\simeq \mathcal{M}_0$) for  $\alpha> \alpha_c$, and otherwise  $\mathcal{M}$ decreases with $K$ as $\mathcal{M}\simeq\mathcal{M}_0 K^{g_\alpha/2}$.

In my analytical approximation,  $f_{\rm v}$ and $y$ are determined using equation~(\ref{eq:fe}) and~(\ref{eq:y}) with $\mathcal{M}=\mathcal{M}_0$. This implies that the variations of $\mathcal{M}$ when $\alpha < \alpha_c$ are neglected. However this remains a reasonable approximation because in this case, as will be shown below, the emission is dominated by the base of the emitting region (i.e. around  $z\sim z_0$).  The resulting values for $f_{\rm v}$ are shown in Fig.~\ref{fig:1}d. It turns out that they agree roughly with the result of the simulations close to the base of the jet. 

The magnetic field can then be estimated as:
\begin{equation}
B=B_0 \eta^{b} K^{\frac{s+\alpha-3}{2}}=B_0\left(\frac{R}{R_0}\right)^{-b},
\end{equation}
with 
\begin{equation}
B_0=\frac{8f_0^{2-\alpha}S_0}{\eta^{b} y\tan{\phi} } \sqrt{\frac{P_J G}{(1+\xi_{e}+\xi_{\rm p})f_{\rm v}(\gamma\beta)^{9}(\gamma-1)|g_\alpha|c^{3}}},
\end{equation}
and
\begin{equation}
R_0=\eta\Delta R_0,
\end{equation}
with
\begin{equation}
\Delta R_0=\frac{ y \gamma^3\beta^3\tan{\phi}c}{4f_0^{1/q}S_0^{1/2}G^{1/2}},
\end{equation}
and $\eta=1+R_b/\Delta R_0$.
In the case $\alpha>\alpha_c$, $b=(4-2\alpha)/(3-\alpha)$ while if $\alpha<\alpha_c$, the magnetic profile becomes independent of $\alpha$ and $b=\gamma_a$.
This scaling holds for $R_1>R>R_0$, where:
\begin{equation}
\frac{R_1}{R_0}=r\simeq\left(\frac{f_0}{f_1}\right)^{1/q}.
\end{equation}
At larger distances (i.e. $ R>R_1$) the magnetic field decays due to adiabatic losses:
\begin{equation}
B=B_0 r^{-b} \left(\frac{R}{R_1}\right)^{-\gamma_a}.
\end{equation}
As can be seen, the magnetic field profile can be approximated as a broken power-law shape. Corresponding to the dissipative and passive regions. 
The specific case $\alpha=\alpha_c$ is important because for $\gamma_a=4/3$, it corresponds to white noise (i.e. $\alpha=0$). In this case the logarithmic dependence on $K$  in equation~(\ref{eq:eps1}) makes the power-law approximation to  the magnetic field profile problematic, because both the collision Mach number $\mathcal{M}$ and magnetic field profile  depend on $\ln K$. In the analytical approximation, I estimate the SED produced by the section of the jet located between the peak of $\tilde{\epsilon}$ at $z_m=z_0 \exp(1/q)$ and $z_f$, by replacing the $\ln K$ term by a constant:
\begin{equation}
\ln \bar{K}= \left[1+\ln\left(f_0/f_1\right)\right]/2,
\end{equation}
 i.e.  the  average of the value at $z_m$ and that at $z_f$. Then, the collision Mach number can be approximated as a constant:
\begin{equation}
\mathcal{M} \simeq \sqrt{\frac{4G}{(\gamma_a-1)\ln {\bar K}}},
\end{equation}
and 
\begin{equation}
B\simeq B_m\left(\frac{R}{R_m}\right)^{-b},
\end{equation}
with
\begin{equation}
R_m=\eta_m z_m \tan{\phi},
\end{equation}
\begin{equation}
\eta_m=1+R_b/(z_m \tan{\phi}),
\end{equation}
and
\begin{equation}
B_m=\frac{8f_0^{2-\alpha}S_0 e^{-b/q}}{\eta_m^{b} y\tan{\phi} } \sqrt{\frac{P_J G \ln{\bar{K}} }{(1+\xi_{e}+\xi_{\rm p})f_{\rm v}(\gamma\beta)^{9}(\gamma-1)c^{3}}}. 
\end{equation}

\subsubsection{Emission properties}\label{sec:emprop}
We can now, use the  power-law approximation for the magnetic field profile to estimate the properties of the SED as described in Appendix~\ref{sec:plbprof}.
There are three distinct regimes for the SED: 
\begin{enumerate}

\item $\alpha<\alpha_c$: This regime is illustrated by the case $\alpha=-1$ in Fig.~\ref{fig:1}. 
The $B$ profile  can then be approximated as a simple power-law with slope $b=\gamma_a$ from $R=R_0$ to $+\infty$.
According to equation~(\ref{eq:tnfreq1}). The break frequency is at:
\begin{equation}
\nu_t=\nu_0\left[-(a+1)\Gamma(a)\right]^\frac{-2}{(p+4)(a+1)},
\label{eq:tnfreq1}
\end{equation}
where $\nu_0$ and $a$ are defined by equations~(\ref{eq:nu0}) and (\ref{eq:a}) respectively.
At shorter frequencies $\nu<< \nu_t$ the jet emits in the partially self-absorbed regime and the SED can be approximated by the power-law of equation~(\ref{eq:fnuinterm}).
The spectral index is given by equation~(\ref{eq:alphab}), for $\gamma_a=4/3$, this reduces to:
\begin{equation}
\alpha_{\rm T}=\frac{2p+13}{4p+18}\simeq 0.65.
\label{eq:alphatpassive}
\end{equation}
Above $\nu_T$ the jet emits in the optically thin regime as given by equation~(\ref{eq:fnuopthin}) i.e. a power-law with spectral index $(1-p)/2$.

\item $\frac{2p+1}{p+2}>\alpha>\alpha_c$: This regime  is illustrated by  the cases $\alpha=0.6$, 1 and 1.2 in Fig.~\ref{fig:1}.
The index of the magnetic field profile is  $b=(4-2\alpha)/(3-\alpha)>6/(p+5)$ and the SED can be approximated as a double broken power-law, with breaks at frequencies $\nu_s$ and $\nu_t$ given by equations (\ref{eq:nus2}) and (\ref{eq:tnfreq1}) respectively.
Below $\nu_s$ the SED is dominated by the emission of the passive outer region of the jet in the partially absorbed regime. The flux is given by equation~(\ref{eq:fnuinterm2}) and the slope of the spectrum is $\alpha_{\rm T}\simeq0.65$  given by equation~(\ref{eq:alphab}). Above $\nu_s$ the dissipative region dominates in the partially absorbed regime as described by equation~(\ref{eq:fnuinterm}). According to equation~(\ref{eq:alphab}) the spectral index is :
\begin{equation}
\alpha_{\rm T}=\frac{(2p+13)(1-\alpha)}{18+4p-\alpha(10+2p)}\label{eq:alphat}
\end{equation}
Above $\nu_T$ the jet emits in the optically thin regime as given by equation~(\ref{eq:fnuopthin}) i.e. a power-law with spectral index $(1-p)/2$.

\item $\frac{2p+1}{p+2}<\alpha$: This regime is illustrated by  the case $\alpha=1.5$ in Fig.~\ref{fig:1}.
 Most of the dissipation occurs at large distance and the emission is dominated at all frequencies by the passive region. 
The break frequency $\nu_t$ is then given by equation~(\ref{eq:nut2}). At low frequencies the emission is in the partially absorbed regime  as given by equation~(\ref{eq:fnuinterm2}). The slope of the spectrum is again given by equation~(\ref{eq:alphab}) resulting in $\alpha_{\rm T}\simeq 0.65$. Above $\nu_T$ the jet emits in the optically thin regime as given by equation~(\ref{eq:fnuopthin}) i.e. a power-law with spectral index $(1-p)/2$.
\end{enumerate}

  \begin{figure}
 \includegraphics[width=\linewidth]{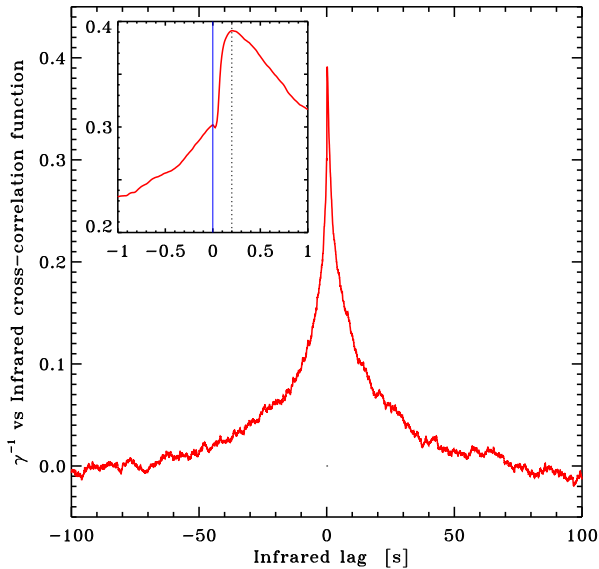}
 \caption{Cross correlation function of 1/$\gamma(t)$ and predicted IR flux from the fiducial  flicker noise simulation with constant jet power. See Section~\ref{sec:variability} for details}
 \label{fig:3}
 \end{figure}

As can be seen in Fig.~\ref{fig:1} those analytical approximations of the SED are in reasonable agreement with the results from the simulations.
Fig.~\ref{fig:1} clearly illustrate the sensitivity of the photon spectrum on the power spectrum of the injected fluctuations. A comparison of panels~\ref{fig:1}a and~\ref{fig:1}e shows that overall, the shape of the SED follows the same trends as that of the PSD.  PSDs with more power at low Fourier frequencies, lead to SEDs with more power radiated at lower photon frequencies. 
 In the case of white noise ($\alpha=0$) the dissipation profile is in agreement with the  previous derivation by Beloborodov (2000). In  this case, most of the dissipation occurs very close to the black hole and then the dissipation rate decreases very quickly (like $z^{-5/3}$).  As a consequence the specific energy profile is steep and therefore the adiabatic losses are not compensated. The photon spectrum is more inverted than in most most of the observed SEDs. 
 
\subsection{Variability}\label{sec:variability}

So far I have discussed only time averaged properties of the jet emission. However, as this emission occurs in the course of a succession of transient events (i.e. the collisions of ejecta), the internal shock model is intrinsically time-dependent. It is an interesting feature of this model that it naturally predicts a strong variability of the jet emission over a broad range of time-scales. Those timing properties will be studied in detail somewhere else. Here I would like to illustrate this feature of the model through a single example that may be relevant to the observations. 
Figure~\ref{fig:2} shows sample light curves and power spectra obtained from the simulation with flicker noise, non-linear jet Lorentz factor fluctuations, constant jet power and fast dissipation. The time average properties obtained for this simulation are shown in Fig.~\ref{fig:linnonlin} and discussed in Section~\ref{sec:flickernoise}. For what regards the variability properties, the jet behaves like a low-pass filter. As the shells of plasma travel down the jet, colliding and  merging with each other, the highest frequency velocity fluctuations are gradually damped and the size of the emitting regions increases. As can be seen in Fig.~\ref{fig:2}, the jet is strongly variable in the optical and IR bands originating primarily from the base of the emitting region and become less and less variable at longer frequencies produced at larger distances from the black hole. 

 In this simulation the amplitude of the variability is large, the frationnal rms amplitude integrated between 0.1 mHz and 50 Hz is 1.08, 0.96, 0.54, 0.40 and 0.20 at $2\times 10^{18}$, $2\times 10^{14}$, $2\times 10^{13}$,$2\times 10^{12}$ and $2\times 10^{10}$ Hz respectively.
The observations also show significant flickering in the Infrared and optical band  (Kanbach et al. 2001; Casella et al. 2010; Gandhi et al. 2010).  At least part of this fast IR/optical variability is likely to arise from the jet, possibly through internal shocks. The observed variability amplitude appears however smaller (10-30 percent) than what I obtain here. This might be due to the presence of a less variable emissions components like the outer accretion disc contributing to the optical/IR observed emission. Nevertheless the shape of the observed IR/optical power spectrum is quite similar to the synthetic one at $2\times10^{14}$ Hz (shown in Fig.~\ref{fig:2}; see Fig 3 of Casella 2010 and Fig. 11 of Gandhi et al. 2010  for comparison)

{It is worth noting that  the delays due to propagation of radiation inside the shell are neglected in the simulations. Since the escaping radiation is at most partially absorbed, the medium can be considered as optically thin. The photon escape time is then  $\tilde{t}_{e}\sim R/c$. Taking into account photon propagation would therefore  smear out the variability on time scales shorter than $~R/c$ and this may have some effects on the simulated light curves. However the shortest observed time-scales are related to the shock dissipation time-scale $\tilde{t}_{c}$ which is given by  equation~(\ref{eq:tc}).  For the parameters considered, the dissipation is significantly longer than the escape time-scale:
\begin{equation}
\frac{\tilde{t}_{c}}{\tilde{t}_{e}}= \frac{x f_{\rm V}}{2 y \gamma\beta \tan{\phi}}\simeq 9.
\end{equation}
This approximation is thus unlikely to have any significant effect on the results presented here.}

Another interesting property of the observed IR variability is the existence of correlations with the fast X-ray variability originating from the accretion flow.  
 In particular, Casella et al. (2010) measured de cross-correlation function of the X-ray and IR light curves and found significant correlations between the two bands with the infrared photons lagging behind the X-rays by about 100 ms. At this stage the present internal shock model does not describe  how the fluctuations of the Lorentz factor may be related to the X-ray fluctuations. Nevertheless, we can make simple guesses. Fig.~\ref{fig:3} shows the results obtained if one assumes that the X-ray flux scales like $1/\gamma(t)$. This case may correspond to a situation in which both the jet and X-ray emission tap their energy from a single energy reservoir and are anti-correlated (Malzac, Merloni \& Fabian 2004). Alternatively, this could also correspond to a situation in which the jet kinetic power is a constant and the Lorentz factor fluctuations are related to fluctuations of the matter density  in the innermost part of the accretion. This would induce fluctuations of the jet mass loading, causing a reduction of the Lorentz factor when the ejected shells are more massive. At the same time, in hot accretion flow models the X-ray flux scales approximately like the square of the accretion flow density, causing the jet Lorentz factor and X-ray flux to be anti-correlated.  In any case, Fig.~\ref{fig:3}, shows a cross correlation function that is very similar to that observed by Casella et al. (2010). As can be seen from the inset of Fig~\ref{fig:3}, the simulated IR light curves lag by about 200 ms behind the X-rays. 
Although a detailed modelling of timing data is devoted to future work, it is encouraging to see that this simple model naturally  predicts variability properties that are close to those observed in real sources. 
 
\section{Conclusions}

In this paper, I have explored the internal shock model for jet emission in the context of X-ray binaries, with an emphasis on the dependence of the time-averaged SED of the jet  on the PSD of the fluctuations of the jet Lorentz factor. 
This exploration is based on both a new semi-analytical formalism,  and a new numerical code describing the hierarchical merging of the ejecta and their associated synchrotron emission.

For conical jets, the observed nearly flat radio to IR SEDs  require electron acceleration and dissipation distributed along the jet over a wide range of distances. I have shown that, in the context of internal shock models, this implies that fluctuation of the jet Lorentz factor have to be injected over a wide range of timescales. If instead, those fluctuations have a narrow range of time scales, as e.g. the sine fluctuations studied in Sec.~\ref{sec:sin}, dissipation occurs mostly around one particular distance determined by  the  time-scale of the fluctuations. Electrons propagating along the jet can be accelerated when they cross this dissipation region, but they are quickly cooled down by adiabatic expansion as they travel away. This leads to an radio-IR SED that is more inverted ($\alpha\simeq0.65$) than usually observed in X-ray binaries.  Obtaining an exactly  flat SED in the radio to IR range requires instead that the PSD of the Lorentz factor fluctuations scales like $1/f$ over a broad range of Fourier frequencies.  A steeper PSD, leads to more  dissipation at large distance from the black hole and a steeper radio-IR SED. On the other hand a flatter PSD leads to dissipation closer to the base of the emitting region and the SED is inverted. For a power-law PSD of injected fluctuations,  I derived simple analytical estimates of the spectral index as a function of the slope of the PSD (see Section~\ref{sec:emprop}).  These analytical formulae are in agreement with the results of the simulations.
I have also shown that the internal shock model predicts a strong, wavelength dependent variability, that is similar to what observed in X-ray binaries. The variability properties of the model remain to be explored in more details and the  detailed modelling of the observed X-ray vs IR fast timing correlations may provide important clue on the dynamics of the coupling between accretion and ejection processes. For instance the preliminary results presented here suggest that on short time scale the jet Lorentz factor is anti-correlated with the X-ray luminosity and this could be due to  fluctuations of the mass accretion rate on short time scales causing simultaneously a variation of the X-ray luminosity and mass loading of the jet. This also suggests that the fluctuations of the jet Lorentz factor might be related to the observed X-ray variability.

In fact, the formation of jets appears to be intrinsically connected to accretion processes and compact jets are usually believed to be launched from  accretion discs (Blandford \& Payne 1982; Ferreira et al. 2006). Even in the case when the jet is ultimately powered by the rotation of the black hole (Blandford Znajek 1977), an accretion flow is required and most likely drives the fluctuations of the jet. The variability of the accretion flow in the direct environment of the compact object (where the jet is also launched) can be traced through X-ray variability studies.  It turns out that the X-ray power spectrum of X-ray binaries is close to, although not exactly, flicker noise. At low Fourier frequencies the power decreases approximatively like $S(f) \propto f^{-1.3}$ while at high frequencies ($>10^{-2}$ Hz) the power spectra often show more complicated structures such as broad Lorentzian or quasi-periodic oscillations (Reig, Kylafis \& Papadakis  2002, 2003; Gilfanov \&Arefiev 2005; Gilfanov 2010). If this connection between the variability of the accretion flow and the fluctuations of the jet Lorentz factor is real,  the model predicts a clear relation  between the X-ray timing properties and the radio to IR spectral properties of X-ray binaries.  Such a connection, which  will be explored  in future works, could also be used to probe the dynamics of accretion-ejection onto a black hole.

\section*{Acknowledgments}
The author thanks the Institute of Astronomy (Cambridge) for hospitality. This work has received fundings from PNHE in France, and from the french Research National Agency: CHAOS project ANR-12-BS05-0009 (http://www.chaos-project.fr). The author thanks the anonymous referee for useful comments that helped improve the manuscript.

\appendix

\section[]{Analytical estimates of the synchrotron emission}\label{sec:analyticem}

\subsection{Jet emission}
The relativistic electrons emit through synchrotron.
For electrons with a power-law energy distribution, the synchrotron emissivity is (see e.g. Rybicki \& Lightman 1979): 
\begin{equation}
j_{\tilde{\nu}}=K_j\xi_e B^{\frac{p+5}{2}}\tilde{\nu}^{-\frac{p-1}{2}},
\label{eq:jnu}
\end{equation}
where
\begin{equation}
K_j=\frac{\sqrt{3}e^{3}i_{\gamma}}{16\pi^2 m_e^2 c^4 (p+1)} \Gamma\left(\frac{3p+19}{12}\right)\Gamma\left(\frac{3p-1}{12}\right)\left(\frac{m_ec}{3 e}\right)^{-\frac{p-1}{2}}.
\label{eq:kj}
\end{equation}
{ where the constant $i_\gamma$ is defined as:
\begin{equation}
i_\gamma^{-1}=\bar{\gamma}_e \int_{\gamma_{\rm min}}^{\gamma_{\rm max}} \gamma^{-p}d\gamma.
\label{eq:igm1} 
\end{equation}
}
This radiation can be self-absorbed. The absorption coefficient is :
\begin{equation}
\alpha_{\tilde{\nu}}=K_{\alpha}\xi_e B^{\frac{p}{2}+3}\tilde{\nu}^{-(p+4)/2},
\label{eq:alphanu}
\end{equation}
with
\begin{equation}
K_\alpha=\frac{\sqrt{3}e^3 i_{\gamma}}{64\pi^2 m_e^3 c^4}\left(\frac{3e}{2\pi m_e c}\right)^{p/2}\Gamma\left(\frac{3p+2}{12}\right)\Gamma\left(\frac{3p+22}{12}\right)
\label{eq:kalpha}
\end{equation}

The specific intensity at the surface of the jet is approximated as:
\begin{equation}
\tilde{I}_{\tilde{\nu}}=\frac{j_{\tilde{\nu}}}{\alpha_{\tilde{\nu}}}(1-e^{-\alpha_{\tilde{\nu}}R}).
\label{eq:inu}
\end{equation}
If the jet is in a steady state in the observer's frame the jet SED is given by:
\begin{equation}
F_{\nu}(\nu=\tilde{\nu}\delta)=\delta^2\frac{\gamma\beta c}{2D^2}\int_0^{+\infty}f_{\rm v}R \tilde{I}_{\tilde{\nu}}(\tilde{\nu})d\tilde{t}.
\end{equation}
where $\delta=\left[\gamma(1-\beta\cos\theta)\right]^{-1}$ is the usual Doppler factor, $\theta$ the angle between the jet axis and the line of sight and $D$ the distance to the source.
In the case of a conical jet:
\begin{equation}
F_{\nu}(\nu=\tilde{\nu}\delta)=\frac{\delta^2}{2D^2\tan{\phi}}\int_{R_0}^{+\infty}f_{\rm v}R \tilde{I}_{\tilde{\nu}}(\tilde{\nu})dR.
\label{eq:fnufi}
\end{equation}

\subsection[]{Estimates of a spectrum for a magnetic field with power-law profile}\label{sec:plbprof}

{I now assume that the magnetic field has a power-law dependence with $R$ in the range $R_0$, $R_f$:
\begin{equation}
B=B_0 (R/R_0)^{-b},
\end{equation}
and I derive estimates for the emission spectrum of this section of the jet. Equation~(\ref{eq:fnufi}) becomes: 
\begin{equation}
F_{\nu}=\frac{\delta^{-\frac{1}{2}}R_0B_0^{-\frac{1}{2}}\nu^{\frac{5}{2}}}{2D^2\tan{\phi}}\frac{K_j}{K_\alpha}\int_{R_0}^{R_f} f_{\rm v} \left(\frac{R}{R_0}\right)^{1+\frac{b}{2}}(1-e^{-\tau_{\nu}})dR,
\end{equation}
with
\begin{equation}
\tau_{\nu}(R)=\tau_{0}\left(R/R_0\right)^{l-bd},
\label{eq:taunu}
\end{equation}
and,
\begin{equation}
\tau_{0}=K_\alpha\xi_e B_0^{d}R_0 (\nu/\delta)^{-\frac{p+4}{2}},
\end{equation}
and 
\begin{equation}
d=3+\frac{p}{2}, \label{eq:dpar}
\end{equation}
and
\begin{equation}
l=1.\label{eq:lpar}
\end{equation}
The synchrotron optical depth at $R_f$ is $\tau_{f}=\tau_{\nu}(R_{f})$.
If we then neglect the variations of the volume filling factor:
\begin{equation}
F_{\nu}=\frac{f_{\rm v}\delta^{-1/2}R_0^2B_0^{-1/2}\nu^{5/2}}{D^2\tan{\phi}\tau_{0}^a}\frac{a}{b+4}\frac{K_j}{K_\alpha}F(a,\tau_{0},\tau_{f}),
\label{eq:fnufa}
\end{equation}
with
\begin{equation}
F(x,y_1,y_2)=\int_{y_1}^{y_2} \tau^{x-1}(1-e^{-\tau})d\tau,
\end{equation}
and 
\begin{equation}
a=\frac{b+4}{2(l-bd)}.
\label{eq:a}
\end{equation}

At the highest frequencies both $\tau_0$ and $\tau_{f}$ are small, this is the optically thin regime in which
\begin{equation}
F(a,\tau_{0},\tau_{f})\simeq-\frac{\tau_0^{a+1}}{a+1}+\frac{\tau_f^{a+1}}{a+1}.
\end{equation}
Then:
\begin{equation}
F_{\nu}=\left[r^{\frac{(b+4)(1+a)}{2a}}-1\right]\frac{\delta^{\frac{p+3}{2}}K_j \xi_e f_{\rm v} R_0^3B_0^{d-\frac{1}{2}}a}{D^2\tan{\phi}(b+4)(1+a)}\nu^{-\frac{p-1}{2}},
\label{eq:fnuopthin}
\end{equation}
where $r=R_f/R_0$.
At low frequencies,  both $\tau_0$ and $\tau_{f}$ are very large, this is the saturated optically thick regime.
\begin{equation}
F(a,\tau_{0},\tau_{f})\simeq-\frac{\tau_0^{a}}{a}+\frac{\tau_f^{a}}{a},
\end{equation}
\begin{equation}
F_{\nu}=\left[r^{\frac{b+4}{2}}-1\right]\frac{\delta^{-1/2}f_{\rm v} R_0^2B_0^{-1/2}}{D^2\tan{\phi}(b+4)}\frac{K_j}{K_\alpha}\nu^{5/2}.
\end{equation}
From equation~(\ref{eq:taunu}) we see that if $b>l/d$ and for sufficiently large  $r$, there is a range of frequencies for which $\tau_0>>1>>\tau_f$.  The condition $b>l/d$ implies that $a<0$. Since $a$ is negative the $F$ function can be approximated as:
\begin{equation}
F(a,\tau_{0},\tau_{f})\simeq\Gamma(a)+\frac{\tau_f^{a+1}}{a+1}.
\label{eq:fbgr}
\end{equation}
If $-1<a<0$ the first term in equation~(\ref{eq:fbgr}) dominates and $F\simeq\Gamma(a)$, while for $a<-1$ (i.e. $b<(2l+4)/(2d-1)$)\footnote{here we assume that $d>1/2$ and $l>0$}, the emission at the highest frequencies is dominated by the optically thin synchrotron produced at $R_f$ and  $F\simeq\tau_f^{a+1}/(a+1)$.

In the case $b<l/d$, there is a range of frequencies at which the optical depth is larger in the outer parts of the jet. For $r>>1$, we have $\tau_f>>1>>\tau_0$.  The $F$ function can then be approximated as:
\begin{equation}
F(a,\tau_{0},\tau_{f})\simeq -\Gamma(a)-\frac{\tau_0^{a+1}}{a+1}+\frac{\tau_f^{a}}{a}.
\label{eq:fbsm}
\end{equation}
Then, if $a<0$  (i.e. $b<-4$) the first term dominates in equation~(\ref{eq:fbsm}) and $F\simeq-\Gamma(a)$, while for $a>0$ (i.e $-4<b<l/d$)  the emission is is dominated by optically thick synchrotron from the outer regions: $F\simeq\tau_f^{a}/a$.

To summarize, for $b>(2l+4)/(2d-1)$ (and  for $b<-4$) there is a range of intermediate frequencies for which it is possible to choose $r>>1$ so that $\tau_f<<1$ and $\tau_0>>1$ (respectively  $\tau_f>>1$ and $\tau_0<<1$ ) and the emission at a given frequency is a mixture of optically thick and thin emission from different regions of the jets. In this partially absorbed regime:
\begin{equation}
F(a,\tau_{0},\tau_{f})\simeq \frac{b}{|b|}\Gamma(a),
\end{equation}
\begin{equation}
F_{\nu}=\frac{b\Gamma(a)\delta^{2-\alpha_{\rm T}}f_{\rm v}R_0^{2-a}B_0^{-ad-\frac{1}{2}}K_ja}{|b|D^2\tan{\phi} (b+4) K_\alpha^{1+a}\xi_e^a}\nu^{\alpha_{\rm T}},
\label{eq:fnuinterm}
\end{equation}
with
\begin{equation}
\alpha_{\rm T}=\frac{5}{2}+\frac{a}{2}(p+4)=\frac{(2p+13)(b-1)}{(p+6)(b-1)+p+4}.
\label{eq:alphab}
\end{equation}
The observed turnover frequency is at the intersection of the optically thin and optically thick  asymptotic branches:
\begin{equation}
\nu_t=\nu_0\left[\frac{r^{\frac{(b+4)(a+1)}{2a}}-1}{(a+1)\Gamma(a)b/|b|}\right]^\frac{2}{(p+4)(a+1)},
\label{eq:tnfreq1}
\end{equation}
where
\begin{equation}
\nu_0=\delta\left(K_{\alpha}\xi_e R_0 B_0^{d}\right)^{\frac{2}{p+4}}.
\label{eq:nu0}
\end{equation}
The observed frequency of transition to the fully absorbed regime is given by: 
\begin{equation}
\nu_s=\nu_0\left[\frac{r^{\frac{b+4}{2}}-1}{a\Gamma(a)b/|b|}\right]^\frac{2}{a(p+4)}.
\label{eq:tnfreq2}
\end{equation}
On the contrary if $-4<b<(2l+4)/(2d-1)$,  the intermediate optically thick regime does not exist because the emission at any frequency is dominated by the region with large radius $R_f$. The spectrum evolves directly from optically thin to saturated optically thick, the transition frequency is at: 
\begin{equation}
\nu_t=\nu_0\left(\frac{a}{a+1}\frac{r^{\frac{(4+b)(1+a)}{2a}}-1}{r^{\frac{4+b}{2}}-1}\right)^{\frac{2}{p+4}}.
\end{equation}

\subsection[]{Effects of the passive outer zone on the appearance of the jet}

In the physical situation  considered in this paper, the energy density does not drop instantly to 0 at $R_f$, due to adiabatic losses the magnetic field decays slowly as $B=B_0r^{-b}(R/R_f)^{-\gamma_a}$. If  $\gamma_a>(2l+4)/(2d-1)$, the partially absorbed regime of the passive zone dominates over the optically thick emission of the dissipative zone at low frequencies:
\begin{equation}
F_{\nu}=\frac{a_\gamma \Gamma(a_\gamma)\delta^{2-\alpha_{\rm \gamma}}f_{\rm v}R_0^{2-a_\gamma}B_0^{-a_\gamma d-\frac{1}{2}}K_j\xi_e^{-a_\gamma}}{D^2\tan{\phi}K_\alpha^{1+a_\gamma}(\gamma_a+4) r^{(a_\gamma-a)(l-bd)}}\nu^{\alpha_{\rm \gamma}}
\label{eq:fnuinterm2}
\end{equation}
with
\begin{equation}
a_\gamma=\frac{\gamma_a+4}{2(l-d\gamma_a)}
\end{equation}
and
\begin{equation}
\alpha_{\rm \gamma}=\frac{5}{2}+\frac{a_{\gamma}}{2}(p+4)=\frac{(2p+13)(\gamma_a-1)}{\gamma_a(p+6)-2}.
\label{eq:alphag}
\end{equation}

When $-4<b<(2l+4)/(2d-1)$ this power-law connects directly with the optically thin emission of the dissipative zone (given by equation~\ref{eq:fnuopthin}). The SED is a broken power-law  with break frequency:
\begin{equation}
\nu_t=\nu_1\left[\frac{a(\gamma_a+4)}{a_\gamma(b+4)}\frac{r^{\frac{4+b}{2a}}-r^{-\frac{4+b}{2}}}{\Gamma(a_\gamma) (1+a) r^{-a_\gamma (l-bd)}}\right]^{\frac{2}{2\alpha_\gamma+(p-1)}},
\label{eq:nut2}
\end{equation}
with 
\begin{equation}
\nu_1=\delta\left( K_\alpha \xi_e  R_0 B_0^{d} \right)^{\frac{1+a_\gamma}{\alpha_\gamma+(p-1)/2}}.
\end{equation}

For $b>(2l+4)/(2d-1)$ the emission from the passive zone connects with the partially absorbed regime of the dissipative zone (equation~\ref{eq:fnuinterm}) at frequency:
\begin{equation}
\nu_s=\nu_2\left[\frac{a\Gamma(a)(\gamma_a+1)}{a_\gamma \Gamma(a_\gamma)(b+1)}r^{(a_\gamma-a)(l-bd)}\right]^{\frac{1}{\alpha_\gamma-\alpha_{\rm T}}},
\label{eq:nus2}
\end{equation}
where
\begin{equation}
\nu_2=\delta\left( K_\alpha \xi_e  R_0 B_0^{d} \right)^{\frac{a_\gamma-a}{\alpha_\gamma-\alpha_{\rm T}}}.
\end{equation}
}
\section[]{Numerical estimate of the emission from an ejecta}\label{sec:numem}

This section presents the simplified treatment used for the calculation of the radiative properties of the jet in the the simulations. 

Observed fluxes are always a time-average over some exposure time.  This exposure may be longer than the dynamical time-scale of an ejecta. In order to compare the predictions of the model with observations, I have to  sum the time averaged flux of all the shells that are active during a simulated exposure.
 Let's say that we want to calculate the time averaged jet emission at an observed frequency $\nu$ between observed times $t_{i}$ and $t_{f}$. I note that due to photon propagation the reception time of the jet radiation does not correspond to the time in the lab frame. The origin of time $t=0$ in the lab frame is defined as the time at which the first shell is launched. The origin of time in the observer's time, $t_r=0$, is then defined as the time at which this event would be observed on Earth. 
  
 In the analytical model the point of radius $R_b$ where the shells are injected (the base of the jet) was set at $z=0$. Here, in order to make the algebra simpler, I will slightly change the notation. I shift the $z$ axis so that the injection point corresponds to $z=z_b=R_b/\tan{\phi}$. The result is that at every point in the jet, its radius is exactly $R= z \tan{\phi}$.  Then, a photon emitted at position $z$ and at time $t$  is received at a time $t_r=t-(z-z_b)\mu/c$, where  $\mu=\cos{i}$ and $i$ is the angle of the line of sight with respect to the jet velocity.

For simplicity, the effects of of shock propagation within an ejecta are neglected. For the purpose of the calculation of its emission an ejecta is considered to be fully homogeneous at all times. Similarly, photon travel delays within an ejecta are neglected and the emission from the surface of an ejecta is uniform at all times in the co-moving frame.  
 The longitudinal extension of the ejecta (and its evolution) must however be determined in order to estimate the evolution of the particle and energy densities that control its luminosity and spectral properties.
 In the code an ejecta is characterised by a constant mass $M$, dissipation rate $\dot \epsilon$, and bulk velocity $\beta$c.  When, in the course of the simulation,  any of these parameters is changed (due e.g. to a collision with another ejecta),  the ejecta is removed and replaced by a new ejecta with up-dated parameters.  Therefore in this scheme a specific ejecta exists only during the limited time during which those defining parameters remain contants. 
 Let us consider an ejecta that is active between times $t_0$ and $t_1$ (measured in the lab frame). It is  localised trough its boundaries at positions $z^{-}$ and $z^{+}$ (with  $z^{+}>z^{-}$). The boundaries travel at (constant)  fractional velocities $\beta^{-}$ and  $\beta^{+}$.  
 The length of the shell is:
 \begin{equation}
 H=z^{+}-z^{-}=H_0+(\beta^{+}-\beta^{-})c (t-t_0)
 \end{equation}
 where $H_0=z^{+}_0-z^{-}_0$ is the length of the shell at $t_0$
Assuming that the ejecta is homogeneous, the centre of momentum is located at a position:
\begin{equation}
z=z_0+\beta c (t-t_0),
\end{equation}
where
\begin{equation}
z_0=\frac{\gamma^{+}z_0^{+}+\gamma^{-}z_0^{-}}{\gamma^{+}+\gamma^{-}},
\end{equation}
and 
\begin{equation}
\beta=(\gamma_{+}\beta_{+}+\gamma_{-}\beta_{-})/(\gamma_{-}+\gamma_{+}).
\end{equation}
The coordinate system of the CM frame is  defined so that the event ($\tilde{t}=0$, $\tilde{z}=0$) corresponds to ($t=t_0$, $z=z_0$) in the lab frame. 
In the CM  frame, the shell boundaries travel at velocities:
\begin{equation}
\tilde{\beta}^{\pm}=\frac{\beta^{\pm}-\beta}{1-\beta\beta^{\pm}},
\end{equation}
by definition $\tilde{\beta}^{+}=-\tilde{\beta}^{-}=\tilde{\beta}$, which can be rewritten as:
\begin{equation}
\tilde{\beta}=c_e \gamma^2(\beta^{+}-\beta^{-})/2,
\end{equation}
where
\begin{equation}
c_e=4\left(2+\frac{\gamma^{+}}{\gamma^{-}}+\frac{\gamma^{-}}{\gamma^{+}}\right)^{-1}.
\end{equation}
The length of the shell in the CM frame is then:
\begin{equation}
\tilde{H}=\tilde{H}_0+{2\tilde{\beta}}c\tilde{t},
\label{eq:hdet}
\end{equation}
with:
\begin{equation}
\tilde{H}_0=c_e\gamma H_0.
\label{eq:h0deh}
\end{equation}
Equations~(\ref{eq:hdet}) and~(\ref{eq:h0deh}), imply that 
$\tilde{H}=c_e\gamma H$.  This differs from the standard formula for length contraction by a factor $c_e$ which corrects for the effects of expansion/contraction of the ejecta. Ignoring $c_e$ would imply that for very different boundary velocities, the expansion/contraction velocity of the shell in CM frame can be larger than the speed of light.

If $\dot{\epsilon}$ is the constant power per unit mass of the shell dissipated through shocks, the evolution of the specific energy of the ejecta is given by:
\begin{equation}
\frac{d\tilde{\epsilon}}{d\tilde{t}}=\dot{\epsilon}-\tilde{\epsilon}(\gamma_a-1) \frac{d\ln{\tilde{V}}}{d\tilde{t}}.
\end{equation}
The formal solution is:
\begin{equation}
\tilde{\epsilon}=\left[\dot\epsilon\int_{0}^{\tilde{t}} \left(\frac{\tilde{V}}{\tilde{V}_0}\right)^{\gamma_a-1} d\tilde{t}+\tilde{\epsilon_0}\right]\left(\frac{\tilde{V}}{\tilde{V}_0}\right)^{1-\gamma_a} ,
\end{equation}
where the subscript $0$ indicate the value at $\tilde{t}=0$.
Assuming that the shell has a cylindrical geometry $\tilde{V}=\pi R^2 \tilde{H}$ and neglecting  the longitudinal expansion losses:
\begin{equation}
\tilde{\epsilon}=\left[\frac{\dot\epsilon z_0}{\gamma \beta c} \frac{x^{2\gamma_a-1}-1}{2\gamma_a-1}+\tilde{\epsilon_0} \right]x^{2-2\gamma_a}, 
\end{equation}
where 
\begin{equation}
x=\frac{R}{R_0}=\frac{z}{z_0}=1+\frac{\gamma \beta c\tilde{t}}{z_0} .
\end{equation}

The evolution of the magnetic field is then:
\begin{equation}
B=\sqrt{\frac{8Mc^2 \tilde{\epsilon} x^{-2} h^{-1}
}{(1+\xi_e+\xi_p)R_0^2 \tilde{H}_0}},
\end{equation}
where
\begin{equation}
 h=\tilde{H}/\tilde{H}_0=1+\frac{2\tilde{\beta} c}{\tilde{H}_0}\tilde{t}=1+\frac{2\tilde{\beta} z_0}{\gamma \beta\tilde{H}_0}( x-1).
 \end{equation}
The instantaneous flux emitted by the shell is approximated as:
\begin{equation}
F_\nu(t)=\delta^{3} \frac{R \tilde{H}}{2D^2} \tilde{I}_{\tilde{\nu}},
\end{equation}
where $\tilde{I}_{\tilde{\nu}}$ is given by equation~(\ref{eq:inu}).

Since the light crossing time of a shell can be longer than the exposure time, we have to take into account the delays due to the shell light crossing time. In fact the received flux at time $t_r$ is
\begin{equation}
F_{s\nu} (t_r)=\int_{t_0}^{t_1} F_\nu(t') g_l(t',t_r) d t'.
\label{eq:fess}
\end{equation}
The green function $g_l$ appearing in equation~(\ref{eq:fess}) is then:
\begin{equation}
g_l(t,t_r)=\frac{H_e(t_r-t^{+})H_e(t^{-}-t_r)}{t^{-}-t^{+}},
\end{equation}
where $t^{+}$ and $t^{-}$ are the times at which the radiation emitted at time $t$ from  $z^{+}$ and $z^{-}$ respectively, is received: 
\begin{equation}
t^{\pm}=t^{\pm}_0+(t-t_0)(1-\mu\beta^{\pm})
\end{equation}
with
\begin{equation}
t^{\pm}_0=t_0-(z_0^{\pm}-z_b)\mu/c
\end{equation}

and  $H_e$ represents the heaviside function defined such as:
$H_e(x)=1$ for $x\ge0$, and  $H_e(x)=0$ otherwise.
The observed flux averaged over an exposure time $\Delta_r= t_f-t_i$ can then be calculated as:
\begin{equation}
\bar{F}_\nu= \int_{t_{i}}^{t_{f}}\frac{F_{s\nu}}{\Delta_r} dt_r=\frac{\delta^3 c_e\gamma c}{2D^2 \mu \Delta_r} \int_{t_0}^{t_1}\ R  \tilde{I}_{\tilde{\nu}} G_l\left(t_{i},t_{f},t\right) dt
\label{eq:fnubart}
\end{equation}
where
\begin{equation}
G_l=\left[\min(t_{f},t^{-})-\max(t_{i},t^{+})\right] H_e(t_{f}-t^{+})H_e(t^{-}-t_{i}).
\end{equation}
Equation~(\ref{eq:fnubart}) can be expressed in terms of the known shell parameters at $t_{0}$:
\begin{equation}
\bar{F}_\nu(\nu)=\bar{F}_{\nu0} \int_{\max(1,x_i^{-})}^{\min(x_1,x_f^{+})} \left(\frac{x^6h}{\tilde{\epsilon}}\right)^{\frac{1}{4}} \left(1-e^{-\tau_{\tilde{\nu}}}\right)  h_d dx,
\label{eq:Fnubar}
\end{equation}
with,
\begin{equation}
\bar{F}_{{\nu}0}=\frac{\sqrt{\delta^{-1}R_0^{5}\nu^{5} }}{2D^2 \tan{\phi} \beta  c\Delta_r}\frac{K_j}{K_\alpha}\left[ \frac{c_e^5 \gamma H_0^{5}(1+\xi_e+\xi_p)}{8 Mc^2 }\right]^{1/4},
\end{equation}
\begin{equation}
\tau_{\tilde\nu}=\tau_{\tilde{\nu}0} x^{-\frac{p+3}{2}}(\tilde{\epsilon}/h)^{\frac{p+5}{4}},
\end{equation}
\begin{equation}
\tau_{\tilde{\nu}0} =K_{\alpha}\xi_e R_0^{-\frac{p+3}{2}} \left[\frac{8Mc^2}{(1+\xi_e+\xi_p) c_e \gamma H_0}\right]^{\frac{p+5}{4}} (\delta \nu)^{\frac{1-p}{2}},
\end{equation}
\begin{equation}
x_{1}=1+\frac{\beta c}{z_0}(t_1-t_0),
\end{equation}
\begin{equation}
x_{f}^{\pm}= 1+\frac{\beta c}{z_0}\frac{ t_f-t_0^{\pm}}{1-\mu\beta^{\pm}},
\end{equation}
\begin{equation}
x_{i}^{\pm}=1+\frac{\beta c}{z_0}\frac{ t_i-t_0^{\pm}}{1-\mu\beta^{\pm}},
\end{equation}
and the function $h_d$ is defined as:
\begin{equation}
h_d=\frac{(t^{-}-t_i)c}{\mu H_0}=\frac{z_0}{\mu \beta H_0} (1-\mu\beta^{-})(x-x_i^-)
\end{equation}
 for $ \min(x_i^{+},x_f^{-})> x>x_i^{-}$,
\begin{equation}
h_d= \frac{(t^{-}-t^+)c}{\mu H_0}= h \quad \mathrm{for} \quad x_f^{-}> x> x_i^{+} , 
\end{equation}
\begin{equation}
h_d=\frac{(t_f-t_i)c}{\mu H_0}=\frac{c\Delta_r}{\mu H_0}  \quad  \mathrm{for} \quad  x_i^{+}>x> x_f^{-},
\end{equation}
\begin{equation}
h_d=\frac{(t_f-t_o^+)c}{\mu H_0}=\frac{z_0 }{\mu \beta H_0}  (1-\mu\beta^+) (x_f^{+}-x)
\end{equation}
for  $x_f^{+} >x>\max(x_f^{-},x_i^{+})$, and $h_d=0$ otherwise.

In the numerical scheme, the emission of each ejecta that is active during the time interval of interest is estimated by integrating numerically equation~(\ref{eq:Fnubar}). If the integrand varies by less than 5 percent during this time interval, $\bar{F}_\nu$ is estimated using a simple trapeze formula. Otherwise, a full Romberg integration procedure is applied. The contribution of all the shells is calculated and then added up in order to compute the total time-averaged flux of the source during the exposure. 
\begin{figure}
\includegraphics[width=\linewidth]{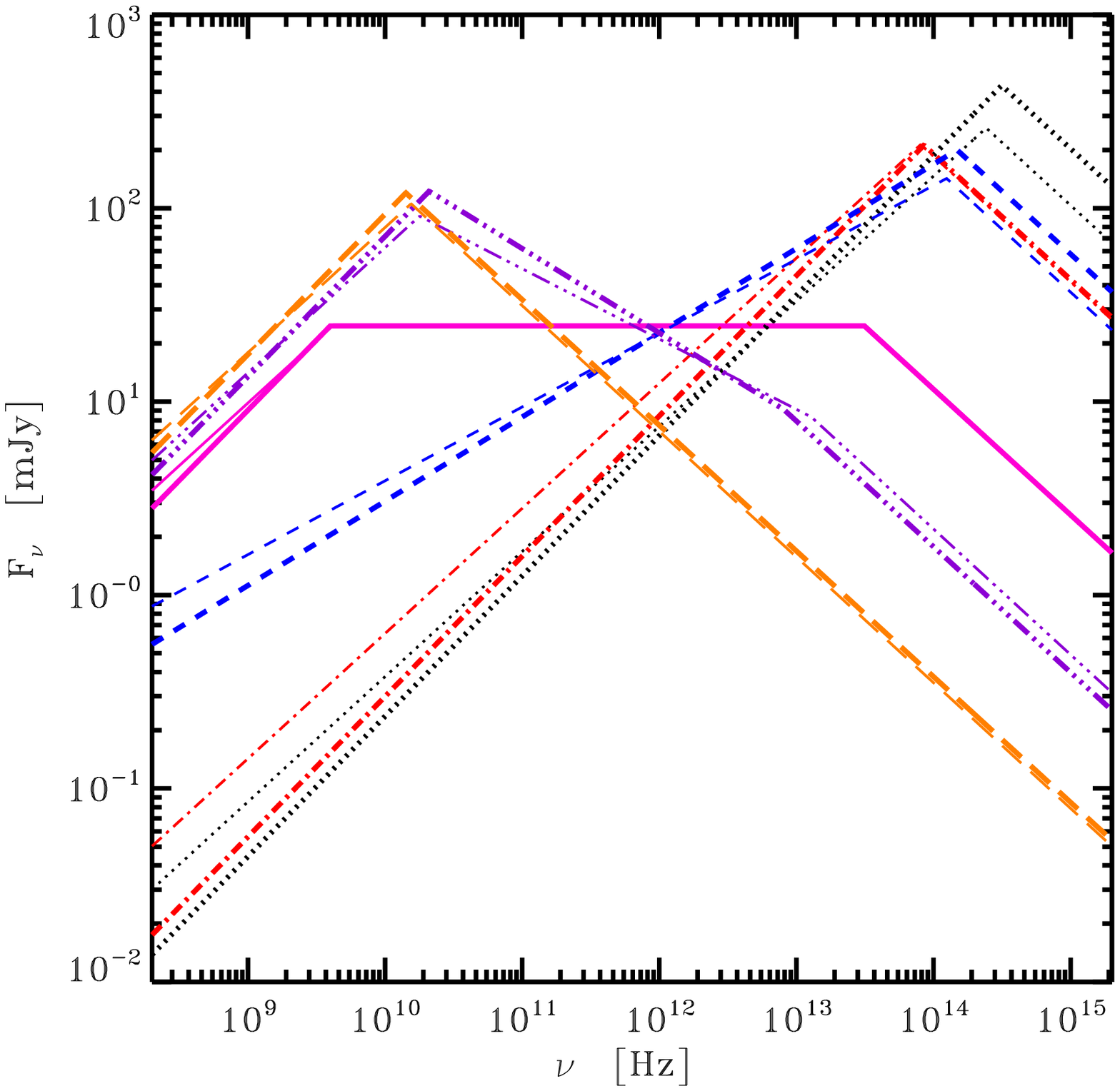}
\caption{Comparison of different analytical estimates of SEDs  predicted by  the internal shock model with a power-law PSD of the Lorentz factor fluctuations ($P(f)\propto f^{-\alpha}$). The thin curves are the same as those shown in Fig~\ref{fig:1}(e) i.e. calculated assuming that the electrons have a constant average energy. The thick curves are for a constant fraction $\chi_a=0.46$ of accelerated electrons. The models parameters and line styles are the same as in Fig~\ref{fig:1}.} \label{fig:8} 
\end{figure}
{
\section{Analytical model with constant number of accelerated particles}\label{sec:cnae}

The details of the evolution of the spectral energy distribution depends not only on the dynamics of the shell collision but also on the microphysics of the acceleration processes.  The latter is both extremely complex and outside the scope of this work.  For this reason, I have used simple prescriptions to deal with the particle kinetics (like for instance the assumption of equipartition). 
In particular, throughout this paper I  assume that the shape of the energy distribution of the accelerated electrons is uniform trough time and along the jet. Only its normalisation varies to account for the changes in internal energy. As a consequence, the number of accelerated electrons per nucleon,  
\begin{equation}
 \chi_a = \frac{m_p\tilde{n}_e}{\tilde{\rho}}=\frac{\xi_e}{1+\xi_e+\xi_p} \frac{m_p\tilde{\epsilon}}{\bar{\gamma}_e m_e c^2},
\end{equation}
is, in general, variable along $z$. It is interesting to note that there is nothing to prevent $\chi_a$ from becoming larger than unity for large values of $\tilde{\epsilon}$ (associated  for example to fluctuations of large amplitude).  In this case the jet would be required to be dominated by electron-positron pairs. Fortunately, this situation never occurs in the examples shown in this paper. Nevertheless, the fraction $\chi_a$ can evolve with $\tilde{\epsilon}$ from a vanishing number of accelerated electrons to almost 100 percent.

 It is not clear whether this is a good prescription. Nevertheless, in order to test the sensitivity of the model to the particle kinetics we can consider a different scenario. In the following, I explore what happens if  we  assume the number of accelerated particles is a constant but their average energy tracks the changes in internal energy.
If the number of accelerated particles is conserved along the jet and neglecting longitudinal expansion, their number density profile $\tilde{n}_e(z)$ depends simply on the geometry of the jet:
\begin{equation}
\tilde{n}_e=\tilde{n}_0(R/R_0)^{-2},
\end{equation}
where
\begin{equation}
\tilde{n}_0=\chi_a \tilde{\rho_0}/m_p,
\end{equation}
is the accelerated particle density at the base of the jet, $\tilde{\rho_0}$ is the mass density given by equation~(\ref{eq:rho}) and taken at $R_0$ and $\chi_a$ is now a constant.
The energy distribution of the accelerated electrons is still a power-law, and I assume that the slope $p$ is a constant, but the average energy of the of the particles can change due to changes in the values of $\gamma_{min}$ and $\gamma_{max}$.  However in the limit where $\gamma_{\rm max} >>\gamma_{\rm min}$, the average Lorentz factor of the accelerated electrons is insensitive to $\gamma_{\rm max}$:
\begin{equation}
\bar{\gamma_e} \simeq \frac{p-1}{p-2}\gamma_{\rm min}
\end{equation}
Then using equation~(\ref{eq:equipart}),  one finds that the integral $i_\gamma^{-1}$, given by  equation~(\ref{eq:igm1}) is not a constant anymore:
\begin{equation}
i_{\gamma}^{-1}\simeq i_{\gamma_0}^{-1} B^{4-2p} \left(\frac{R}{R_0}\right)^{4-2p}
\label{eq:igamma}
\end{equation}
with
\begin{equation}
i_{\gamma_0}^{-1}=\frac{(p-2)^{1-p}}{(p-1)^{2-p}} \left(\frac{\xi_e}{8\pi \tilde{n}_0 m_e c^2}\right)^{2-p}
\label{eq:igamma0}
\end{equation}

Then by injecting the expression of $i_{\gamma}$ given by equation (\ref{eq:igamma}) in equations ~(\ref{eq:jnu}),~(\ref{eq:kj}),~(\ref{eq:alphanu}), and ~(\ref{eq:kalpha}),  the dependence of  the jet emissivity and absorption coefficients on $B$ is changed compared to our fiducial model. This dependence is now: 
\begin{equation}
j_{\tilde{\nu}}=K_{j0}\xi_e B^{\frac{5p-3}{2}}\left(\frac{R}{R_0}\right)^{2p-4} \tilde{\nu}^{-\frac{p-1}{2}},
\end{equation}
\begin{equation}
\alpha_{\tilde{\nu}}=K_{\alpha0}\xi_e B^{\frac{5p}{2}-1}\left(\frac{R}{R_0}\right)^{2p-4} \tilde{\nu}^{-(p+4)/2},
\end{equation}
where the constants $K_{j0}$  and $K_{\alpha0}$ are  given by equations~(\ref{eq:kj}) and~(\ref{eq:kalpha}) respectively  with $i_\gamma$ replaced by $i_{\gamma0}$ given in equation~(\ref{eq:igamma0}).

This can be used to compute the jet emission exactly in the same way as before and as described in Appendix~\ref{sec:analyticem} with the scaling of the synchrotron optical depth, i.e. the $l$ and $d$ parameters given by equations~(\ref{eq:dpar}) and ~(\ref{eq:lpar}), are changed to $l=2p-3$ and $d=5p/2-1$.  The slope of the partially self-absorbed region, given by equation~(\ref{eq:alphab}), becomes:
\begin{equation}
\alpha_{\rm T}=\frac{(12p-7)(b-1)}{(5p-2)(b-1)+p+4}
\label{eq:alphaTconstn}
\end{equation}
We see that the dependence of $\alpha_{\rm T}$ is quantitatively different from that of the model with constant $\bar{\gamma_e}$, but for typical values of $p\sim$2--3 this difference is actually small. 
Also, the cases $b=1$ corresponding in our model to flicker noise power spectrum of Lorentz factor fluctuations still corresponds to a flat spectrum. In absence of dissipation (i.e. $b=\gamma_a=4/3$) the slope of the spectrum  ($\alpha_{\rm T}\simeq0.72$) is then only slightly more inverted than in the model used in the rest of the paper ($\alpha_{\rm T}\simeq0.65$).

In the framework of the internal shock model, and for a power-law PSD of the jet Lorentz factor fluctuations, the dependence of the SED on $\alpha$ is  changed only slightly. Depending on the value of $\alpha$, the  three spectral  regimes discussed in section~\ref{sec:emprop}  are still present: 
\begin{enumerate}
\item $\alpha<\alpha_c$: the SED can be approximated as a broken power-law with break frequency given by equation~(\ref{eq:tnfreq1}).
At $\nu<< \nu_t$ the jet emits in the partially self-absorbed regime and the SED can be approximated by the power-law of equation~(\ref{eq:fnuinterm}).
The spectral index is given by equation~(\ref{eq:alphaTconstn}) with  $b=\gamma_a=4/3$, this reduces to:
\begin{equation}
\alpha_{\rm T}=\frac{12p-7}{8p+10}\simeq 0.72.
\label{eq:alphatpassive2}
\end{equation}
Above $\nu_T$ the jet emits in the optically thin regime as given by equation~(\ref{eq:fnuopthin}) i.e. a power-law with spectral index $(1-p)/2$.

\item $\frac{4p-3}{3p-2}>\alpha>\alpha_c$: 
The index of the magnetic field profile is  $b=(4-2\alpha)/(3-\alpha)> (2l+4)/(2d-1)$ and the SED can be approximated as a double broken power-law, with breaks at frequencies $\nu_s$ and $\nu_t$ given by equations (\ref{eq:nus2}) and (\ref{eq:tnfreq1}) respectively.
Below $\nu_s$  the flux is given by equation~(\ref{eq:fnuinterm2}) and the slope of the spectrum is $\alpha_{\rm T}\simeq0.72$  given by equation~(\ref{eq:alphaTconstn}). Above $\nu_s$,  according to equation~(\ref{eq:alphaTconstn}) the spectral index is :
\begin{equation}
\alpha_{\rm T}=\frac{(12p-7)(1-\alpha)}{10+8p-\alpha(2+6p)}\label{eq:alphatnconst}
\end{equation}
Above $\nu_T$ the jet emits in the optically thin regime as given by equation~(\ref{eq:fnuopthin}).

\item $\frac{4p-3}{3p-2}<\alpha$: The SED is a broken power-law. 
The break frequency $\nu_t$ is given by equation~(\ref{eq:nut2}). At low frequencies the emission is  given by equation~(\ref{eq:fnuinterm2}). The slope of the spectrum is again given by equation~(\ref{eq:alphaTconstn}) resulting in $\alpha_{\rm T}\simeq 0.72$. Above $\nu_T$ the jet emits in the optically thin regime as given by equation~(\ref{eq:fnuopthin}).
\end{enumerate}

Fig.~\ref{fig:8} presents a comparison between the analytical estimates of the SED obtained assuming a constant $\bar{\gamma}_e$  and those assuming a constant fraction of accelerated electrons for the models discussed in section~\ref{sec:powerlawpsd} and displayed in Fig.~\ref{fig:1}.  The value of $\chi_a$ was set to  $\chi_a=0.46$ in order to match that obtained in the constant $\bar{\gamma}_e$ model with $\alpha=1$.   Overall the difference between the two sets of SEDs are relatively weak. This suggests that the results  are not very sensitive to approximate treatment of the microphysics. Nevertheless, it is worth noting that in this approximation the SED is calculated under the assumption that for each photon  frequency $\nu$ of interest, there are  accelerated electrons radiating most of their synchrotron radiation around $\nu$. The fact that our electron distribution is spread only between $\gamma_{\rm min}$ and $\gamma_{\rm max}$ actually introduces additional breaks in the SED. In the model with constant $\bar{\gamma}_e$, the choices of $\gamma_{\rm min}=1$ and $\gamma_{\rm max}=10^6$ ensure that those breaks occur outside of the radio-IR range. In the constant $\chi_a$ model, which assumes changes in $\gamma_{\rm min}$ and $\gamma_{\rm max}$, this is not granted and may introduce some spectral complexity that is not taken into account in the present estimates.

}

\bsp

\label{lastpage}


\begin{thebibliography}{99}

\bibitem[\protect\citeauthoryear{Beloborodov}{2000}]{2000ApJ...539L..25B} 
Beloborodov A.~M., 2000, ApJ, 539, L25 

\bibitem[\protect\citeauthoryear{Blandford 
\& Znajek}{1977}]{1977MNRAS.179..433B} Blandford R.~D., Znajek R.~L., 1977, MNRAS, 179, 433 



\bibitem[\protect\citeauthoryear{Blandford {\ 
K&ouml}nigl}{1979}]{1979ApJ...232...34B} Blandford R.~D., K{\"o}nigl A., 1979, ApJ, 232, 34 

\bibitem[\protect\citeauthoryear{Blandford 
\& McKee}{1976}]{1976PhFl...19.1130B} Blandford R.~D., McKee C.~F., 1976, PhFl, 19, 1130 

\bibitem[\protect\citeauthoryear{Blandford 
\& Payne}{1982}]{1982MNRAS.199..883B} Blandford R.~D., Payne D.~G., 1982, MNRAS, 199, 883 

\bibitem[\protect\citeauthoryear{B{\"o}ttcher 
\& Dermer}{2010}]{2010ApJ...711..445B} B{\"o}ttcher M., Dermer C.~D., 2010, ApJ, 711, 445 

\bibitem[\protect\citeauthoryear{de 
Bruyn}{1976}]{1976A&A....52..439D} de Bruyn A.~G., 1976, A\&A, 52, 439 

\bibitem[\protect\citeauthoryear{Casella et 
al.}{2010}]{2010MNRAS.404L..21C} Casella P., et al., 2010, MNRAS, 404, L21 


\bibitem[\protect\citeauthoryear{Chaty, Dubus, 
\& Raichoor}{2011}]{2011A&A...529A...3C} Chaty S., Dubus G., Raichoor A., 2011, A\&A, 529, A3


\bibitem[\protect\citeauthoryear{Condon 
\& Dressel}{1973}]{1973ApL....15..203C} Condon J.~J., Dressel L.~L., 1973, ApL, 15, 203 


\bibitem[\protect\citeauthoryear{Corbel et al.}{2000}]{2000A&A...359..251C} Corbel S., Fender R.~P., Tzioumis A.~K., Nowak M., McIntyre V., Durouchoux P., Sood R., 2000, A\&A, 359, 251 

\bibitem[\protect\citeauthoryear{Corbel 
\& Fender}{2002}]{2002ApJ...573L..35C} Corbel S., Fender R.~P., 2002, ApJ, 573, L35 


\bibitem[\protect\citeauthoryear{Daigne 
\& Mochkovitch}{1998}]{1998MNRAS.296..275D} Daigne F., Mochkovitch R., 1998, MNRAS, 296, 275 



\bibitem[\protect\citeauthoryear{Done et al.}{1992}]{1992ApJ...400..138D} 
Done C., Madejski G.~M., Mushotzky R.~F., Turner T.~J., Koyama K., Kunieda 
H., 1992, ApJ, 400, 138 

\bibitem[\protect\citeauthoryear{Falcke}{1996}]{1996ApJ...464L..67F} Falcke H., 1996, ApJ, 464, L67 

\bibitem[\protect\citeauthoryear{Falcke 
\& Markoff}{2000}]{2000A&A...362..113F} Falcke H., Markoff S., 2000, A\&A, 362, 113 

\bibitem[\protect\citeauthoryear{Fender et al.}{2000}]{2000MNRAS.312..853F} 
Fender R.~P., Pooley G.~G., Durouchoux P., Tilanus R.~P.~J., Brocksopp C., 
2000, MNRAS, 312, 853 

\bibitem[\protect\citeauthoryear{Fender}{2001}]{2001MNRAS.322...31F} Fender 
R.~P., 2001, MNRAS, 322, 31 

\bibitem[\protect\citeauthoryear{Ferreira et al.}{2006}]{2006A&A...447..813F} Ferreira J., Petrucci P.-O., Henri G., Saug{\'e} L., Pelletier G., 2006, A\&A, 447, 813 

\bibitem[\protect\citeauthoryear{Gandhi et al.}{2010}]{2010MNRAS.407.2166G} 
Gandhi P., et al., 2010, MNRAS, 407, 2166 

\bibitem[\protect\citeauthoryear{Ghisellini, Maraschi, 
\& Treves}{1985}]{1985A&A...146..204G} Ghisellini G., Maraschi L., Treves A., 1985, A\&A, 146, 204 

\bibitem[\protect\citeauthoryear{Gilfanov}{2010}]{2010LNP...794...17G} 
Gilfanov M., 2010, LNP, 794, 17

\bibitem[\protect\citeauthoryear{Gilfanov 
\& Arefiev}{2005}]{2005astro.ph..1215G} Gilfanov M., Arefiev V., 2005, astro, arXiv:astro-ph/0501215 

\bibitem[\protect\citeauthoryear{Hjellming 
\& Johnston}{1988}]{1988ApJ...328..600H} Hjellming R.~M., Johnston K.~J., 1988, ApJ, 328, 600 

\bibitem[\protect\citeauthoryear{Jamil, Fender, 
\& Kaiser}{2010}]{2010MNRAS.401..394J} Jamil O., Fender R.~P., Kaiser C.~R., 2010, MNRAS, 401, 394 

\bibitem[\protect\citeauthoryear{Jamil {\ 
B&ouml}ttcher}{2012}]{2012ApJ...759...45J} Jamil O., B{\"o}ttcher M., 2012, ApJ, 759, 45 

\bibitem[\protect\citeauthoryear{Joshi {\ 
B&ouml}ttcher}{2011}]{2011ApJ...727...21J} Joshi M., B{\"o}ttcher M., 2011, ApJ, 727, 21 

\bibitem[\protect\citeauthoryear{Kaiser, Sunyaev, 
\& Spruit}{2000}]{2000A&A...356..975K} Kaiser C.~R., Sunyaev R., Spruit H.~C., 2000, A\&A, 356, 975 

\bibitem[\protect\citeauthoryear{Kaiser}{2006}]{2006MNRAS.367.1083K} Kaiser 
C.~R., 2006, MNRAS, 367, 1083 

\bibitem[\protect\citeauthoryear{Kanbach et 
al.}{2001}]{2001Natur.414..180K} Kanbach G., Straubmeier C., Spruit H.~C., 
Belloni T., 2001, Natur, 414, 180 

\bibitem[\protect\citeauthoryear{Konigl}{1981}]{1981ApJ...243..700K} Konigl 
A., 1981, ApJ, 243, 700 


\bibitem[\protect\citeauthoryear{van der Klis}{2004}]{2004astro.ph.10551V} 
van der Klis M., 2006, in Compact stellar X-ray sources, Lewin \& van der Klis (eds), Cambridge University Press, p 39,  arXiv:astro-ph/0410551 

\bibitem[\protect\citeauthoryear{Lyubarsky}{2010}]{2010MNRAS.402..353L} 
Lyubarsky Y.~E., 2010, MNRAS, 402, 353 


\bibitem[\protect\citeauthoryear{Malzac}{2013}]{2013MNRAS.429L..20M} Malzac 
J., 2013, MNRAS, 429, L20 

\bibitem[\protect\citeauthoryear{Malzac, Merloni, 
\& Fabian}{2004}]{2004MNRAS.351..253M} Malzac J., Merloni A., Fabian A.~C., 2004, MNRAS, 351, 253 


\bibitem[\protect\citeauthoryear{Marscher}{1977}]{1977ApJ...216..244M} 
Marscher A.~P., 1977, ApJ, 216, 244 

\bibitem[\protect\citeauthoryear{Marscher}{1980}]{1980ApJ...235..386M} 
Marscher A.~P., 1980, ApJ, 235, 386 

\bibitem[\protect\citeauthoryear{Miller-Jones, Fender, 
\& Nakar}{2006}]{2006MNRAS.367.1432M} Miller-Jones J.~C.~A., Fender R.~P., Nakar E., 2006, MNRAS, 367, 1432 

\bibitem[\protect\citeauthoryear{Perucho, Bosch-Ramon, 
\& Khangulyan}{2010}]{2010A&A...512L...4P} Perucho M., Bosch-Ramon V., Khangulyan D., 2010, A\&A, 512, L4 

\bibitem[\protect\citeauthoryear{Rees}{1978}]{1978MNRAS.184P..61R} Rees 
M.~J., 1978, MNRAS, 184, 61P 


\bibitem[\protect\citeauthoryear{Rees 
\& Meszaros}{1994}]{1994ApJ...430L..93R} Rees M.~J., Meszaros P., 1994, ApJ, 430, L93 


\bibitem[\protect\citeauthoryear{Reig, Papadakis, 
\& Kylafis}{2002}]{2002A&A...383..202R} Reig P., Papadakis I., Kylafis N.~D., 2002, A\&A, 383, 202 


\bibitem[\protect\citeauthoryear{Reig, Papadakis, 
\& Kylafis}{2003}]{2003A&A...398.1103R} Reig P., Papadakis I., Kylafis N.~D., 2003, A\&A, 398, 1103 


\bibitem[]{}Rybicki G. B., Lightman A. P., 1979, Radiative processes in
astrophysics, ed. Rybicki, G. B. \& Lightman, A. P. (Wiley- Interscience)


\bibitem[\protect\citeauthoryear{Spada et al.}{2001}]{2001MNRAS.325.1559S} 
Spada M., Ghisellini G., Lazzati D., Celotti A., 2001, MNRAS, 325, 1559 


\bibitem[\protect\citeauthoryear{Timmer 
\& Koenig}{1995}]{1995A&A...300..707T} Timmer J., Koenig M., 1995, A\&A, 300, 707 

\bibitem[\protect\citeauthoryear{Uttley, McHardy, 
\& Vaughan}{2005}]{2005MNRAS.359..345U} Uttley P., McHardy I.~M., Vaughan S., 2005, MNRAS, 359, 345 

\bibitem[\protect\citeauthoryear{Zdziarski, Pjanka, 
\& Sikora}{2013}]{2013arXiv1307.1309Z} Zdziarski A.~A., Pjanka P., Sikora M., 2013, MNRAS, submitted, arXiv, arXiv:1307.1309 



\end{thebibliography}
\end{document}